\documentclass{article}
\usepackage{amsmath}
\usepackage{amsfonts}
\numberwithin{equation}{section}
\usepackage[hidelinks]{hyperref}
\usepackage{secdot}

\usepackage[numbers,compress]{natbib}

\def\p{\partial}
\def\O{\mathcal{O}}
\def\C{\mathcal{C}}
\def\a{\alpha}
\def\b{\beta}
\def\g{\gamma}
\def\G{\Gamma}
\def\r{\rightarrow}
\def\F{\mathcal{F}}

\def\d{\delta}

\def\e{\epsilon}

\newcommand{\be}{\begin{equation}}
\newcommand{\ee}{\end{equation}}
\newcommand{\bea}{\begin{eqnarray}}
\newcommand{\eea}{\end{eqnarray}}
\newcommand{\bi}{\begin{itemize}}
\newcommand{\ei}{\end{itemize}}

\title{ On correlation functions in $J\bar{T}$-deformed CFTs
\vspace{3mm}}

\author{
Monica Guica \vspace{2mm}\\
\\\vspace{2mm}
\emph{\normalsize Institut de Physique Th\'eorique, CEA Saclay, CNRS},
\emph{91191 Gif-sur-Yvette, France}\\\vspace{2mm}
\emph{\normalsize  Department of Physics, Stockholm University,
AlbaNova, 106 91 Stockholm, Sweden}
\\ 
\emph{\normalsize Nordita, Roslagstullsbacken 23, SE-106 91 Stockholm, Sweden}
}
\date{}

\begin{document}

\maketitle

\begin{abstract}
\vspace{0.3cm}

\noindent The $J\bar T$ deformation, built from the components of the stress tensor and of a $U(1)$ current, is a universal irrelevant deformation of two-dimensional CFTs that preserves the left-moving conformal symmetry, while breaking locality on the right-moving side. 
Operators in the $J\bar T$-deformed CFT are  naturally labeled by the left-moving position and  right-moving momentum and transform in representations of the one-dimensional extended conformal group.
We derive an all-orders formula for the spectrum of conformal dimensions and charges of the deformed CFT, which we cross-check at leading order using conformal perturbation theory. We also compute the linear corrections to the one-dimensional OPE coefficients and comment on the extent to which the correlation functions in $J\bar T$-deformed CFTs can be obtained from field-dependent coordinate transformations.
  
\flushright\emph{Dedicated to the memory of P.G.O Freund 
}  
  
\end{abstract}

\tableofcontents

\section{Introduction}

There has been plenty of recent interest in deformations of two-dimensional QFTs by  irrelevant operators constructed from bilinears of conserved currents \cite{smzam}, of which the so-called $T\bar T$ deformation  \cite{smzam,Cavaglia:2016oda} is the best studied example \cite{Dubovsky:2012wk,Caselle:2013dra,Dubovsky:2017cnj,
Cardy:2018sdv,Cardy:2018jho,Aharony:2018vux,Datta:2018thy,Aharony:2018bad,Chen:2018eqk,Conti:2018tca,Bonelli:2018kik,Lashkevich:2018jmo,Chang:2018dge,Baggio:2018rpv}. Remarkably, this deformation is solvable in a certain sense and the deformed spectrum is determined from that of the original QFT through a universal formula.  $T\bar T$ deformed CFTs have also found several very interesting applications in string theory \cite{Giveon:2017nie,Giveon:2017myj,Asrat:2017tzd,Giribet:2017imm,Araujo:2018rho} and in holography \cite{McGough:2016lol,Shyam:2017znq,Kraus:2018xrn,Cottrell:2018skz,Gorbenko:2018oov,Shyam:2018sro,Hartman:2018tkw,Taylor:2018xcy}.

Another remarkable property of $T\bar T$ deformed QFTs is that they are believed to be UV complete, albeit non-local. The reason for this claim is that certain on-shell quantities, such as the S-matrix,   appears to be well-defined up to  arbitrarily high energies, at least for one sign of the deformation parameter \cite{Dubovsky:2012wk,Dubovsky:2012sh,Dubovsky:2013ira}. The status of off-shell quantities such as correlation functions is less clear but, given the proposal that the $T\bar T$ deformation produces a theory of quantum gravity in two dimensions \cite{Dubovsky:2017cnj,Dubovsky:2018bmo}, one should perhaps not expect  local observables to exist beyond perturbation theory.

It is interesting to ask about the ultraviolet behaviour of other deformations in the Smirnov-Zamolodchikov class \cite{smzam} and to find an appropriate set of local or non-local observables. In this article, we will address this question for the case of the $J\bar T$ deformation of two-dimensional CFTs\footnote{Note that the $J\bar T$ deformation can be defined for any two-dimensional QFT that possesses a $U(1)$ current. However, in this more general case, it is not known whether the deformed spectrum is universal. Also, the left-conformal invariance, which is essential for the analysis of this paper, will not be present if it is not present in the initial QFT.  } \cite{Guica:2017lia}, which is  comparable to $T\bar T$ in what regards its universality, but  different from it in an interesting way and possibly more  tractable.  The $J\bar T$ deformation is defined by adding to the  action a composite operator constructed from a chiral $U(1)$ current $J$ and the right-moving stress tensor $\bar T$

\be
\frac{\p }{\p \mu} S(\mu) =  \int d^2 z \, (J \bar T)_\mu
\ee
where the subscript indicates that the deformation is always performed using the current and stress tensor of the deformed theory. Just like $T\bar T$, the  $J\bar T$ deformation is solvable. The finite-size spectrum was first derived in \cite{Guica:2017lia} for the case of vanishing chiral anomaly and in section $6$ of \cite{kutasov} for the general case. The holographic dictionary for  $J\bar T$ deformed CFTs was derived in \cite{cssint} and the string-theoretical realization of a single-trace version of $J\bar T$ was studied in \cite{kutasov,Apolo:2018qpq}. The modular  properties of the partition function of $J\bar T$ deformed CFTs were studied in \cite{Aharony:2018ics}.  Other  related work is \cite{Nakayama:2018ujt}.

At the CFT point, the $J\bar T$ operator has dimension $(1,2)$, so it is marginal from the point of view of the left conformal symmetry and irrelevant from the point of view of the right one.  As a consequence of adding this operator to the CFT action, the deformed theory becomes non-local on the right, but still remains local and conformal on the left, at least at leading order in conformal perturbation theory. 
Given this structure, it is natural to  label operators 
by their left-moving position $z$ and right-moving momentum $\bar p$. Assuming the $SL(2,\mathbb{R})_L $ symmetry is unbroken at higher orders in the deformation parameter,  we can view the $J\bar T$-deformed CFT as a one-dimensional CFT - obtained by  dimensional reduction along the null direction  $\bar z$  - where operators come in continuous towers labeled by $\bar p$.
Since the left-moving global symmetries $SL(2,\mathbb{R})_L \times U(1)_J$ admit  the usual infinite-dimensional extension, these operators will transform in Virasoro$_L \times U(1)$ Ka\v{c}-Moody representations. 


Thus, a natural set of observables in $J\bar T$-deformed CFTs are the correlation functions of the operators $\O_{i, \bar p}(z)$. These one-dimensional  correlation functions are determined in terms of the usual  conformal data: operator dimensions, charges, and the OPE coefficients

\be
h_i (\mu \bar p) \;\;\;\;\;\;\;\;\;\; q_i(\mu \bar p)\;\;\;\;\;\;\;\;\;\; \C_{ijk} (\mu \bar p_l) \label{confdata}
\ee
Due to the irrelevant deformation,  these data can now depend on the dimensionless combination $\mu \bar p$,
where $\mu$ is the deformation parameter, a null vector with dimension of $(length)^{-1}$. This structure is virtually identical with that previously studied in the context of  non-relativistic holography \cite{Guica:2010sw};
the main difference between $J\bar T$ and the deforming operator  in those works is that $J\bar T$ is double-trace and has universal correlation functions with the operators in the undeformed CFT.  As a consequence, the above conformal data are related to the corresponding data in the original CFT$_2$ in a universal way. 

Note that from the point of view of the one-dimensional conformal data \eqref{confdata}, the combination  $\mu \bar p$ need not be small;  indeed, below we present exact expressions for $h_i(\mu \bar p)$ and $q_i(\mu\bar p)$ that have no obvious pathology for large $\mu \bar p$. Thus, $J\bar T$-deformed CFTs appear to be well-defined well above the scale set by the irrelevant operator, providing yet another potential example of a two-dimensional QFT that is UV complete without possessing a usual UV fixed point. Even though the non-locality of $J\bar T$-deformed CFTs makes it natural to use the momentum space representation for the right-movers, note that the same amount of data is contained in the correlation functions of the $\O_{i,\bar p}(z)$ operators as in the original position space representation; by contrast,  in  $T\bar T$ deformed QFTs only on-shell observables are believed to be well-defined.  Thus, the $J\bar T$ deformation leads to a slightly different notion of
  asymptotic fragility \cite{Dubovsky:2013ira} than $T\bar T$. The ultraviolet behaviour of $J\bar T$-deformed CFTs is most likely to resemble that of  lightlike dipole theories, a set of non-local field theories based on a star product \cite{Bergman:2000cw,Alishahiha:2003ru,Dasgupta:2001zu}.

The conformal data \eqref{confdata} can in principle be computed  up to any desired order in the deformation using conformal perturbation theory. 
Two  questions of interest are whether: i) we can provide closed-form expressions for these data that are exact in the $\mu \bar p$ parameter and ii) whether the resulting expressions hint towards the existence of an additional structure that may allow one to specify all correlation functions in the deformed CFT in terms of those in the original CFT, and possibly also provide a geometric interpretation of the correlators directly in right-moving position space. That such an interpretation may exist is suggested by the holographic analysis of \cite{cssint},  who found that the expectation value of the stress tensor  was related by a field-dependent coordinate transformation to the expectation value in the undeformed CFT. Similar field-dependent coordinate transformations appear ubiquitously in the $T\bar T$ context \cite{Dubovsky:2017cnj,Dubovsky:2018bmo}, and it would be very interesting to understand their meaning at a precise quantum level.


In this article, we partially answer i) in the affirmative, by providing an exact expression for the deformed conformal dimensions and charges

\be
h(\mu) = h + \frac{ \mu q }{2\pi}\, \bar p +\frac{ \mu^2  k}{16 \pi^2} \,\bar p^2 \;, \;\;\;\;\;\; q(\mu) = q + \frac{\mu k}{4\pi}\, \bar p \label{defspec}
\ee
where $h,q$ represent the left conformal dimension and  charge in the undeformed CFT and $k$ is the coefficient of the chiral anomaly. 
These expressions are obtained  
by applying an infinite boost to the known spectrum of $J\bar T$-deformed CFTs on the cylinder and, as we show, they agree perfectly with leading order conformal perturbation theory calculations. 
%
 We also show that the pattern of  operator mixing implied by the above formula is consistent with the  structure we obtain using conformal perturbation theory. Notice that the shifts in $h,q$ are reminiscent of spectral flow, where the spectral flow parameter depends on the right-moving momentum.



In order for $ii)$ to hold, the OPE coefficients in \eqref{confdata} would also need to take very special values. To check whether this occurs,  we 
 %
 compute the linearized corrections to the three-point functions, using Euclidean conformal perturbation theory. While the structure  we obtain for the deformed correlator is in perfect agreement with the one dictated by one-dimensional conformal invariance, the corrections to the OPE oefficients not only do not vanish, but in fact are nonzero for structural reasons.
 It is not clear to us whether this is an artifact of the particular regularization procedure we have used, which may or may not be suited for  $J\bar T$-deformed CFTs. We leave this interesting question to future work. 
 


 The organisation of this paper is as follows. In section \ref{genstr}, we discuss the general structure of the correlation functions,  
  how to use conformal perturbation theory to  compute the deformed correlators and to what extent the latter can be obtained by applying  field-dependent coordinate transformations to correlation functions in the original CFT. In section \ref{specboost}, we derive the formula for the spectrum of conformal dimensions and charges from the known spectrum  of $J\bar T$-deformed CFTs on the cylinder. In  section \ref{CPT}, we use conformal perturbation theory to check \eqref{defspec}
  to leading order, compute the linearized corrections to two sample three-point functions  and discuss  operator mixing.  We end in section \ref{disc} with a discussion and future directions. Various useful formulae for the correlators, certain integrals and Fourier transforms are relegated to the appendices.

\section{General structure of the correlation functions \label{genstr}}

As explained in the introduction,  $J\bar T$-deformed CFTs are expected to be local and conformal on the left, but non-local on the right. The most natural way of labeling operators in such a theory is in terms of  the left-moving coordinate $z$ (as local operators transform nicely under the conformal group) and the right-moving momentum $\bar p$. In the original CFT, these operators are given by the Fourier transform\footnote{For the purposes of the Fourier transform, the coordinate $\bar z$ is assumed to be real and independent from $z$. } of the local operators with respect to the right-moving coordinate $\bar z$

\be
\O(z,\bar p) = \int d\bar z \, \O(z,\bar z) \, e^{-i \bar p \bar z} \label{obarp}
\ee
Rather than  $\O(z,\bar p)$, we will use the  notation  $\O_{\bar p} (z)$, thus viewing the Fourier-transformed operators as an infinite tower of local operators in a CFT$_1$, which carry a continuous\footnote{It may also be possible to discretize the $\bar p$ label by considering the DLCQ of $J\bar T$-deformed CFTs. } $\bar p$ label. This representation is also natural if we view the $J\bar T$-deformed CFT as a non-relativistic CFT in zero spatial dimensions, with $\bar p$ playing the role of a continuous particle number label. 

As we turn on the deformation, the operators $\O_{\bar p} (z)$ can acquire an anomalous left conformal dimension. Since, as discussed in \cite{cssint,Guica:2010sw}, the deformation parameter is a null vector with dimension of length, the general form of the left dimension of $\O_{\bar p}$ will be $h (\mu \bar p)$. The  correlation functions of the operators $\O_{i, \bar p_i}$  have the form of those in a one-dimensional CFT, but with the conformal data being in principle arbitrary functions of $\mu \bar p_i$. In the following, we review the expected general structure of the correlation functions that follows from one-dimensional conformal symmetry on the one hand and conformal perturbation theory on the other. Apart from  the issue of operator mixing, this section mostly reviews 
 the structure previously discussed  in the context  of non-relativistic CFTs \cite{Guica:2010sw}.

The 
consequences of the $SL(2,\mathbb{R})_L$  invariance at the level of two and three point functions are reviewed in section \ref{constrconf}. In section \ref{cptsetup}, we briefly review how the deformed correlators are computed using conformal perturbation theory, and  
 discuss issues related to operator degeneracy, mixing and how to choose a diagonal basis of operators. In subsection  \ref{fielddepcootr}, we briefly address the question  whether the correlators in $J \bar T$-deformed CFTs may have a simple structure  directly in right-moving position space, and in particular whether this structure can be interpreted in terms of field-dependent coordinate transformations.

\subsection{Constraints from one-dimensional conformal invariance \label{constrconf}}

In this section, we review the well-known form of correlation functions in one dimensional CFTs, as dictated by $SL(2,\mathbb{R})_L$ invariance. The operators we consider carry two additional labels, the right-moving momentum $\bar p$ and
the $U(1)_L$ charge $q(\mu\bar p)$,  which are constrained by  momentum and, respectively, charge conservation. The only additional structure  we require is that when $\mu=0$, the correlators reduce to the Fourier transform of the original CFT$_2$ correlators with respect to the right-moving coordinate, and that the full answer be a perturbative series in $\mu \bar p$. 

In view of these constraints, the two-point functions take the general form

\be
\langle \O_{i,\bar p_1,q_1} (z_1) \O_{j,\bar p_2,q_2}(z_2 ) \rangle = \frac{\mathcal{N}_{ij}(\mu, \bar p)}{z_{12}^{2h(\mu \bar p)}} \, \d(\bar p_1 + \bar p_2)\, \d_{q_1, -q_2}
 \;, \;\;\;\;\;\;\; \bar p \equiv \bar p_1 =- \bar p_2 \label{2pf}
\ee
The coefficients $\mathcal{N}_{ij}$ vanish if the conformal dimensions of $\O_{i}$ and $\O_j$ are different. Throughout this article, we will choose the operator basis and normalization such that

\be 
 \mathcal{N}_{ij} (\mu, \bar p)= \mathcal{N}_i ( \bar p) \, \d_{ij} \label{2pfnorm}
\ee
where  $ \mathcal{N}_i ( \bar p)  $ is the Fourier transform of the right-moving part of the CFT two-point function with respect to $\bar z$ 
\be
 \mathcal{N}_i ( \bar p) = \int_{-\infty}^\infty \frac{d\bar z \, e^{-i\bar p \bar z}}{\bar z^{2\bar h}} =   \frac{ 2\pi (-i)^{2\bar h} \, |\bar p|^{2\bar h-1}}{\Gamma(2\bar h)} \Theta (\bar p) \label{explnorm}
\ee
and $\bar h$ is the right-moving dimension of the operator $\O_i$. This result agrees with previous calculations of Wightman functions in non-relativistic CFTs \cite{Barnes:2010ev}. 

Once the basis has been normalized, the  non-trivial dynamical data of  $J\bar T$ deformed CFTs lies at the level of the three-point functions 

\be
\langle \O_{i,\bar p_1} (z_1) \O_{j,\bar p_2}(z_2 ) \O_{k,\bar p_3} (z_3) \rangle = \frac{\mathcal{C}_{ijk}( \bar p_l, \mu) }{z_{12}^{h_{ij;k}(\mu \bar p_l)} z_{13}^{h_{ik;j}(\mu \bar p_l)} z_{23}^{h_{jk;i}(\mu \bar p_l)}}  \, \, \d(\bar p_1+\bar p_2+\bar p_3) \label{3pf}
\ee
where 
\be
h_{ij;k}(\mu \bar p_l) = h_i (\mu \bar p_1) +  h_j (\mu \bar p_2)-h_k (\mu \bar p_3)
\ee
and the remaining two combinations are given by permutations. We  have dropped for simplicity the charge labels, keeping in mind that the total charge must be zero.   The coefficient $\C_{ijk} (\bar p_l,\mu)$  takes the general form
\be
\C_{ijk} (\bar p_l, \mu) = K_{ijk} (\bar p_l) \, C_{ijk} (\mu \bar p_l) 
\ee
where $K_{ijk} (\bar p_l) $ is a kinematic factor equal to the Fourier transform of the right-moving part of the original CFT$_2$ three-point function and $C_{ijk} (\mu \bar p_l)$  represents the corrected  OPE coefficient, i.e.

\be
C_{ijk}(\mu\bar p_l) = C_{ijk}^{CFT} + \O(\mu \bar p_l) \label{corrOPE}
\ee
where $C_{ijk}^{CFT}$ is the OPE coefficient in the two-dimensional CFT (a number). 
The function $K_{ijjk}(\bar p_l)$  is defined as  
\be
K_{ijk} (\bar p_l) = \int \prod_{k=1}^3 dz_k \frac{e^{-i \bar p_k \bar z_k}}{\bar z_{12}^a \bar z_{23}^b \bar z_{13}^c} \equiv  K (a,b,c;\bar p_l)\, \d(\bar p_1 + \bar p_2 + \bar p_3) 
\ee
where  $a,b,c$ are given by

\be
a = \bar h_i+ \bar h_j-\bar h_k\;, \;\;\;\;\;\;b = \bar h_j+ \bar h_k-\bar h_i\;, \;\;\;\;\;\;c = \bar h_i+ \bar h_k-\bar h_j \label{abc}
\ee
The momentum-conserving delta function ensures that $K(a,b,c;\bar p_l)$ only depends on two of the three momenta. The Fourier transform is performed in appendix \ref{fourier} and yields
\be
K (a,b,c;\bar p_1,\bar p_2) =\frac{4\pi^2 (-i)^{a+b+c}   }{\Gamma(a+c)\Gamma(b)} \, \bar p_1^{a+c-1} \bar p_2^{b-1} \, {}_2 F_1 \left(1-b,a,c+a,-\frac{\bar p_1}{\bar p_2}\right) \label{Kabc}
\ee
where we have chosen to represent $K$ in terms of $\bar p_1$ and $\bar p_2$, both of which are assumed to be positive. Note that for $b$ a strictly positive integer, the hypergeometric function becomes a polynomial of degree $b-1$. We can of course choose to write $K(a,b,c;\bar p_l)$ in terms of any two of the momenta, 
using the conservation equation.  As discussed in appendix \ref{fourier},   depending on the signs of the momenta, the closed-form expression for $K$ may change slightly. Throughout this article, we will exclusively use the representation above in terms of $\bar p_{1,2}>0$, and consequently drop the $\bar p_l$ argument from $K(a,b,c)$.

The non-trivial data that we are interested in computing is $h_i(\mu \bar p)$ and $C_{ijk} (\mu\bar p_l)$.  Higher-point correlation functions can in principle be built   by repeated application of the OPE. These correlators satisfy bootstrap relations, which impose constraints on the  $\mu \bar p$ - dependent dimensions and  OPE coefficients that should be obeyed order by order in the deformation parameter $\mu$.

\subsection{Building the correlators using conformal perturbation theory \label{cptsetup}}

In this section, we review the  standard procedure  to construct correlation functions in the deformed CFT in terms of correlators in the undeformed one. This procedure will be used in section \ref{CPT} to compute various two and three-point functions of interest. 

Even though the $J\bar T$ deformation is defined in terms of the instantaneous $U(1)$ current and stress tensor, let us for now pretend, for simplicity, that the deforming operator is just the $J\bar T$ operator in the undeformed CFT. This assumption does not affect the low-order calculations we perform in this paper; we comment on the differences from the actual Smirnov-Zamolodchikov $J\bar T$ deformation at the end of this section. The correlation functions in such a $J\bar T$ deformed CFT are defined in terms of the original correlators via

\be
\langle \O_1 \O_2 \ldots \O_n \rangle_{\mu_E} = \langle \O_1 \O_2 \ldots \O_n \, e^{\, \mu_E \int \! \! J \bar T} \rangle_{CFT} \label{cptjto}
\ee
where $\mu_E$ is related to the coefficient $\mu$ appearing in the action as

\be
\mu_E = \frac{i \mu}{4\pi^2} \label{muemu}
\ee
and we are using the conventions\footnote{ In these conventions, $
\Delta S_L = - \mu \int dx^+ dx^- \, J_+  T_{--}$.
Passing to Euclidean signature, we have
$$
i\Delta S_L \r - i \mu \int d^2 z \, J_z T_{\bar z \bar z} = \mu_E \int d^2 z\, J \bar T = - \Delta S_E 
$$
where the relation between $T_{\bar z \bar z}$ and $J_z$ used in \cite{cssint} and the usual $\bar T$, $J$ used in the CFT literature is
$
T_{\bar z \bar z} = - \frac{ \bar T}{2\pi} \,, \; J_z = \frac{J}{2\pi}
$. 
The Euclidean measure differs by a factor of $i$ from the conventions of \cite{cssint}. The relation \eqref{muemu} follows from the above. 
%
}  of \cite{Guica:2017lia,cssint}.

To compute the deformed correlation functions to any order in the perturbation parameter $\mu_E$, one simply  needs to bring down powers of $J \bar T$ from the exponent, evaluate the corresponding correlators in the original CFT and then integrate. We denote the corresponding $n^{th}$ order correction as $\d_n$:

\be
\d_n \langle \O_1 \O_2 \ldots \O_n \rangle_{\mu_E} \equiv \frac{\mu_E^n}{n!} \left\langle \O_1 \O_2 \ldots \O_n \left( \int \! \! J \bar T \right)^n \right\rangle_{CFT}
\ee
Since the correlation functions of any number of $J$ and $\bar T$ insertions with the $\O_i$  are determined by the Ward identities, the integrands are given by a tractable formula at any order in perturbation theory, which depends on the original CFT data in a universal way. 

The next step is to perform the integrals, which are in general divergent.  In this paper, we perform the integrals in Euclidean position space and use dimensional regularization to keep track of the divergent terms. Since the physical data of the 
deformed CFT depend on $\bar p$, we subsequently perform a  Fourier transform with respect to the right-moving coordinate  to rewrite the operators in the $(z,\bar p)$ basis. We can then absorb the divergences by defining the renormalized operators

\be
\O_{i, \bar p}^{ren} (z) 
 = \O_{i , \bar p}(z) + \sum_{n=1}^\infty c_n \mu^n (i \bar p)^n \O_{i, \bar p} (z)
\ee
The divergent part of the $c_n$ is chosen to cancel the divergences,  while  their finite part is scheme-dependent and  can be  fixed by choosing the normalization of the  two-point function, for example \eqref{2pfnorm}.  
 Once the $c_n$ have been fixed, the correlation functions of the renormalized operators are finite and  meaningful to any order in conformal perturbation theory. The  anomalous dimensions can be read off as usual from  the coefficients of the $\ln z$ pieces in the two-point function.

\subsubsection*{Building a diagonal operator basis}

As is well known, operators that have the same conformal dimension can mix in conformal perturbation theory. In the following, we would like to discuss operator degeneracy in $J\bar T$-deformed CFTs in the $(z,\bar p)$ representation,  how degenerate operators mix in conformal perturbation theory and how to obtain a diagonal basis of the form \eqref{2pfnorm} in the deformed theory. For simplicity, we assume that the spectrum of conformal dimensions of the original CFT is non-degenerate, except for the unavoidable degeneracies associated with the extended Virasoro$_L \times$ Virasoro$_R \times U(1)_L$ Ka\v{c}-Moody  symmetry.  

Let us start by discussing the original CFT spectrum in the $(z,\bar p)$ representation. The Fourier transform \eqref{obarp} will  mix  an $SL(2,\mathbb{R})_R$ primary with its descendants, but will not mix different $SL(2,\mathbb{R})_R$ representations with each other. This implies that for each Virasoro$_L \times SL(2,\mathbb{R})_R$ primary in the original CFT, we will obtain one Virasoro$_L$ primary in the $(z,\bar p)$ representation\footnote{These primaries will  further assemble into left-moving Virasoro - Ka\v{c}-Moody representations, but we will mostly ignore the additional affine $U(1)$ structure.}. Since the original CFT spectrum is usually specified in terms of Virasoro$_R$ representations, we should first decompose these representations  into  a tower  of global conformal  primaries of the schematic form\footnote{In general, there will be more than one $SL(2,\mathbb{R})_R$ primary at each level, e.g. at level $6$ there is also an operator of the form $(\bar T \bar \p^2 \bar T \O)$. The list \eqref{otower} is supposed to include all primary operators, even though our notation does not fully reflect this.  }

\be
\O \;\;\;\;\;\; (\bar T \O)\;\;\;\;\;\;(\bar T^2 \O) \ldots \label{otower}
\ee
each of which will give rise to a continuous set of operators labeled by the momentum $\bar p$. All these operators  have the same left-moving  dimension $h$. Thus, we find that in the original CFT in the $(z,\bar p)$ representation, there is an infinite number of degenerate operators for each left conformal dimension, which can in principle mix in conformal perturbation theory. 

To understand their mixing, note first that it can only occur between 
operators with the same $\bar p$ quantum number. Furthermore, since  the correlation functions of the deforming operator  are obtained from the original correlators by a simple application of the Ward identities, two operators whose dimensions are initially different will not mix with each other at any order in conformal perturbation theory. Thus, we only need to study the mixing of the operators in the tower \eqref{otower} among themselves, at fixed $\bar p$. 

We now make use of the formula \eqref{defspec} for the exact deformed spectrum, which will be derived in the next section. This formula shows that while the initial degeneracy between operators of the same  $h$ and different $\bar p$ is lifted, the degeneracy between the infinite number of operators appearing in the tower \eqref{otower} at fixed $\bar p$ is not. We would now like to understand how  to build a diagonal basis for these operators order by order in conformal perturbation theory.

If we choose the original CFT operators in \eqref{otower} to be $SL(2,\mathbb{R})_R$ primaries, then the initial operator basis is diagonal 
\be
\langle \O_i \O_j^\dag \rangle = \d_{ij}
\ee
where we have absorbed for simplicity the normalizations \eqref{explnorm} into the individual operators and dropped the $\bar p$ and $q$ labels. To first order the deformation, the corrections to the two-point function take the form
\be
\d_1 \langle \O_i \O_j^\dag \rangle =\mu  M_{ij} +  \mu N_{ij} \ln z \label{mnmat}
\ee
where $M^{}, N^{}$ are hermitean matrices. To rediagonalize the basis, we define the new operators

\be
 \O'_i = S_{ij} \O_j = (S_0 + \mu S_1 + \dots )_{ij} \O_j
\ee
which should satisfy 
\be
\langle \O'_i  {\O'}_j^{\dag} \rangle = 
 (1- 2 \mu \g_i \ln z ) \d_{ij}
\ee
The $\g_i$ denote the linearized corrections to the anomalous dimensions, which can be packaged into a diagonal matrix $\Gamma$. From this, we find that the first few $S_n$ should satisfy

\be
S_0  S_0^\dag = I \;, \;\;\;\;\;\; S_0 M S_0^\dag + S_1  S_0^\dag + S_0 S_1^\dag   =0 \;, \;\;\;\;\; S_0 N S_0^\dag = - 2 \Gamma
\ee
The first relation tells us that $S_0$ is a unitary matrix. If the perturbation did break the initial degeneracy, then the last equation would fix $S_0$ to be some particular unitary matrix, built from the eigenvectors of the matrix $N$. However, since we know that all anomalous dimensions will be the same, $\Gamma \propto I$, from which it follows that\footnote{This implies in particular that  there should be no logarhitmic divergences in the off-diagonal correlators, a fact that we will check explicitly in several examples in section \ref{CPT}.}  $N \propto I$. Thus,  unlike in the case of non-degenerate perturbation theory, we are still free to choose  any  $S_0$. A natural choice is $S_0 = I$, from which there follows that by choosing
\be
 (S_1 + S_1^\dag) = - M
\ee 
we can make the basis diagonal to this order. Since this equation only fixes the sum of $S_1$ and $S_1^\dag$, we are free to choose these matrices to be lower diagonal, for example. It is easy to convince oneself that the same choice can be made at higher  orders in conformal perturbation theory. 

 Taking the operator basis to be $\O_{i,\bar p} = \{ \O_{\bar p}, (\bar T \O)_{\bar p}, (\bar T^2 \O)_{\bar p} \ldots \}$, this implies  that we can always reach a basis of operators  $\O'_{i,\bar p}$  that have diagonal correlators via a transformation of the form

\be
(\bar T^n \O)'_{\bar p}  = (\bar T^n \O)_{\bar p}  + \sum_{k=0}^n\sum_{m=0}^\infty  c_{m,k} \mu^{m} \bar p^{m+2k} (\bar T^{n-k} \O)_{\bar p} \label{lowerdiag}
\ee
i.e. in order to diagonalize the deformed two-point functions of an operator that was a level $2n$ descendant from the point of view of the original Virasoro$_R$ symmetry, we only need to compute its mixing with descendants of equal or lower level. This makes the process of basis diagonalization for these operators iteratively tractable.

So far, we have only discussed how the various Virasoro$_L$ primaries mix with each other. However, in the original CFT the Virasoro$_L$ symmetry was accompanied by an affine $U(1)_L$ symmetry. When the original Virasoro $\times$ Virasoro $\times $ Ka\v{c}-Moody blocks are decomposed with respect to Virasoro$_L \times SL(2,\mathbb{R})_R$ representations, they give rise to a double tower of operators of the schematic form

\be
\begin{array}{cccc}
\O & \;\;\;\;\;\;  \bar T \O & \;\;\;\;\;\;  \bar T^2 \O & \ldots \vspace{1mm} \\
J\O  & \;\;\;\;\;\; \bar T J \O& \;\;\;\;\;\;\bar T^2 J \O & \ldots \vspace{1mm} \\
J^2\O  & \;\;\;\;\;\; \bar T J^2 \O& \;\;\;\;\;\;\bar T^2 J^2 \O &\ldots \\
\vdots  & \;\;\;\;\;\; \vdots & \;\;\;\;\;\;\vdots  \\
\end{array}
\ee
The operators on the second row have the holomorphic dimension increased by one and the same charge as the ones on the first row, while operators on the third row have the holomorphic dimension increased by two and the same charge, etc. In general, the operators on the lower rows mix 
  with holomorphic derivatives of the rows above. Since their charges are the same, all the operators in the above table aquire 
   the same anomalous dimension  at fixed $\bar p$. Consequently, it seems rather clear that  after the deformation the columns  will again combine into representations of the  Virasoro-Ka\v{c}-Moody algebra, which is the full left-moving symmetry group that is expected to survive. 


\subsubsection*{Remarks about the deforming operator}

At the beginning of this section, we made the simplifying assumption that the deforming operator is the  $J\bar T$ operator of the undeformed CFT to all others in the deformation. We would now like to make a few comments on the relation between this deformation and the one introduced by \cite{smzam}. While the discussion to follow does not concern the leading order conformal perturbation theory calculations that we perform in section \ref{CPT}, these issues will eventually need to be addressed at higher orders in perturbation theory.


The Smirnov-Zamolodchikov deformation built from a $U(1)$  current and the right-moving stress tensor is defined as

\be
(J\bar T)_{SZ} \equiv \left( J \bar T - \bar J \Theta \right)_\mu \label{smzam}
\ee
where $\Theta = T_{z\bar z}$ and all the currents are defined in the deformed theory. 
Our first simplifying assumption with respect to \eqref{smzam} is that $J$ is chiral, which means that $\bar J =0$ as an operator   and thus the last term  can be dropped. 
In particular, we imagine that the original CFT has a purely chiral spectrum of charges, which is a rather strong assumption\footnote{ It may also be possible to start from a generic CFT with a non-chiral spectrum and only use the chiral component of the current, which is separately conserved, to define the deforming operator. }
We  additionally assume that there are no contributions from  contact terms associated to the $\bar J$ term that could yield finite results upon integration.

The current that stays chiral along the flow differs from the original CFT current 
%
%
at second order in conformal perturbation theory\footnote{The relation between  the conventions of \cite{kutasov} and ours is $\mu_{there}= - 2 \mu_E$. We also  reinstated the factors of $k$.} \cite{kutasov} 
\be
J\r   J' =   J - 2 \pi^2  \mu_E^2 \, k : \!J \bar T\!: + \ldots \label{jredef}
\ee
Note this correction is not of the form \eqref{lowerdiag}; 
 as we explain in section \ref{opmix}, the reason this does not follow the general pattern  we have argued for is that the current is 
 %
  a purely holomorphic operator. The same argument implies that higher order corrections of the form $\mu^{2m} :\! J \bar T^m\!:$ may also be necessary. The corrections to this operator, which  is expected to stay holomorphic to all orders in perturbation theory, can be systematically computed using the method we  exemplify in section \ref{opmix}.


On the other hand, the form of the perturbative corrections to  the antiholomorphic  stress tensor is somewhat unusual. In \cite{kutasov},  the correction to $\p \bar T$ that follows from the OPE with the deforming operator was computed to linear order in $\mu$, with the result
%
%

 
\be 
 \p \bar T = - 2 \pi \mu_E  J \bar \p \bar T
 \ee
The solution for $\bar T$ is, formally
\be
\bar T \r \bar T'= \bar T +2\pi  \mu_E \int \!\! J \; \bar \p \bar T + \ldots \label{nonloccorr}
\ee
which resembles the expansion to linear order of an operator of the form $\bar T (\bar z +2 \pi \mu_E \int J)$.  Similar   expressions were obtained in the examples of \cite{Guica:2017lia} and in  holography \cite{cssint}. The problem with this correction is that it is non-local from the point of view of the left-movers\footnote{It may be possible to interpret this correction as the perturbative expansion of the  vertex operator $e^{i q_T \varphi}$, where $\varphi$ is the chiral boson associated with the current $J$ via $k \,\p \varphi = J$ and $q_T = \mu k \bar p/4\pi$ is the charge of the stress tensor \eqref{hqtbar} in the deformed theory. We have not investigated the implications of such an interpretation. }, so it does not particularly fit into our current framework, which  keeps locality on the left-moving side manifest.
%
 Fortunately, this non-local correction  
 does not affect the integrated operator; indeed, the integral picks out the $\bar p =0$ component of the operator  \eqref{smzam}, but since the current is chiral, this translates into the $\bar p=0$ component of the corrected stress tensor, for which the correction \eqref{nonloccorr} vanishes. 

Let us remark that, in principle, one can discuss two different notions of $\bar T$ operator in the deformed theory. One operator corresponds to the $\bar z$ component of the Noether current associated with right-moving translations, $\bar T_{Noether}$, which in general becomes non-local when the deformation is turned on. 
This is the operator that  is  suppposed to   appear in \eqref{smzam}. It is not clear how to define this operator abstractly, i.e. without making reference to a Lagrangian.
  The second option is to consider the left primary operator that starts out  as $\bar T_{\bar p}$ in the undeformed CFT, i.e. as the Fourier transform of the right-moving stress tensor, and then is deformed, while preserving the primary condition. As we show in section \ref{CPT}, $\bar T_{\bar p}$  
behaves just like any other operator in the theory; in particular, it receives an anomalous dimension and charge given by \eqref{hqtbar} and it mixes according to \eqref{lowerdiag}.
The relation between $\bar T_{\bar p}$ and $\bar T_{Noether}$  is rather unclear, except at the level of the zero mode.  Indeed, the zero mode $\bar T_{\bar p=0}$ does not acquire a left anomalous dimension and is not corrected in perturbation thory, and thus could still be playing the role of right-moving global translations generator. 
 

To end this discussion, let us point out that there also exists a second notion of 
 $J\bar T$ operator, which is different from that proposed in \cite{smzam}. Rather than first constructing the $J$ and $\bar T_{Noether}$ operators in the deformed theory and then using the OPE to define the composite operator $J\bar T$, one can instead ask for the left primary operator that is the deformation of the double-trace operator $J\bar T$ in the original CFT. As we explained earlier in this section, this operator is given in terms of the original CFT ones
by a relation of the form \eqref{lowerdiag}. When integrated, this $J\bar T$ operator differs from the integral of \eqref{smzam}  at $\O(\mu^2)$, but it equals the integrated $J\bar T$ opearator of the undeformed CFT to all orders in $\mu$, as can be seen from the fact that all corrections to it vanish when   $\bar p =0$. It is this $J\bar T$ operator whose correlation functions we will compute in section \ref{CPT}. 

The correct identification of the deforming operator is clearly very important in obtaining a UV-complete theory\footnote{Assuming that our conjecture about the UV-completeness of $J\bar T$-deformed CFTs is correct.}. One relatively simple test that the deforming operator should pass is whether it can reproduce  the prediction  \eqref{defspec} for the conformal dimensions and  charges. This should provide a non-trivial check of the deforming operator already  at cubic order in $\mu$.

%
%



 \subsection{Relation to field-dependent coordinate transformations \label{fielddepcootr}}
 
So far, in using the $(z,\bar p)$ basis for labeling operators, we have only assumed the minimum amount of structure of $J\bar T$-deformed CFTs, which follows from the symmetries of the problem.  
 The holographic analysis of   \cite{cssint} suggests however that there may be additional structure to the correlation functions  that can be seen directly in (right-moving) position space.  Concretely, \cite{cssint} found that the holographic
  one-point functions of the stress tensor were related to their CFT counterparts via a simple field-dependent coordinate transformation of the form
\be
x^+ \r x^+ \;, \;\;\;\;\;\; x^- \r x^- -  \mu \int^{x^+} \!\! dx'^+ \,\langle J_+(x'^+) \rangle \label{fdepcootr}
\ee
where the integrand is the classical expectation value of $J_+$ in the heavy state dual to a classical background and $x^\pm$ - which correspond to $z,\bar z$ - are the Lorentzian lightlike coordinates. Very similar operator-dependent coordinate transformations, where the integrals are performed over null lines, have been extensively used in the $T\bar T$ context 
 \cite{Dubovsky:2017cnj}.

 One can therefore ask  whether 
a relation of the form 
\be
\langle \O_1(z_1,\bar z_1) \ldots \O_n (z_n,\bar z_n)\rangle_\mu  \stackrel{\mbox{\large{$?$}}}{=}\langle \O_1 (z_1,\bar z_1- \a \,\mu\! \int^{z_1}\!\!\! dz_1'\, J(z'_1)) \ldots \O_n (z_n,\bar z_n-\a\, \mu\! \int^{z_n}\!\!\! dz_n'\, J(z'_n))\rangle_{CFT} \label{transfcorr}
\ee
 holds, for some precise interpretation of the field-dependent coordinate transformation ($\a$ is a numerical factor depending on the current normalization).  There are many possibilities \emph{a priori} for what this interpretation could be: the integral could be over the operator $J$ or over its classical expectation value sourced at the various insertions, the integration contour could be in the complex $z$ plane or along a Lorentzian null line, etc.  The goal of this subsection is to use
our knowledge of correlation functions in $J\bar T$ deformed CFTs to check 
 %
   %
   whether they admit an interpretation in terms of field-dependent coordinate transformations and, if so, how exactly these should be defined. We will restrict our attention to two and three-point functions, which are fixed in terms of the  conformal dimensions \eqref{defspec} and the OPE coefficients.


 Consider first the two-point function, with normalization given by \eqref{explnorm}
\be
\langle \O_{\bar p}(z_1) \O^\dag_{-\bar p} (z_2) \rangle_\mu = \frac{\mathcal{N}(\bar p)}{z_{12}^{2h(\mu)}} = \frac{\mathcal{N}(\bar p)}{z_{12}^{2h}} \, e^{-  \frac{\mu q}{\pi} \bar p \ln z_{12} - \frac{\mu^2 k}{8\pi^2} \bar p^2 \ln z_{12}} \label{2pftest}
\ee
where we have used the explicit formula \eqref{defspec} for the deformed dimensions. 
 Let us first assume, for simplicity, that $k=0$. Fourier transforming both sides with respect to $\bar z_i$, we  obtain

\be
\langle \O(z_1,\bar z_1) \O^\dag (z_2,\bar z_2) \rangle_\mu = \frac{1}{z_{12}^{2h} (\bar z_{12}+ i \mu q/\pi \ln z_{12})^{2\bar h}} \label{2pfpossp}
\ee
The shift in $\bar z$ can be interpreted as a field dependent coordinate transformation of the form \eqref{transfcorr}

\be
\bar z \r \bar z -\frac{i \mu}{2\pi}  \int^z \!\! dz' J(z')
\ee
where $J$ is a classical field sourced at the locations of the two operators, i.e.  it satisfies

\be
\bar \p  J(z)  = 2\pi \sum_i q_i \d(z-z_i) =\bar \p \left(\frac{q}{z-z_1} - \frac{q}{z-z_2} \right)
\ee
If we now 
 compute the $J$ integral with a short distance cutoff at $z_1 + \d$ and $z_2 - \d$, we find exactly the shift in  \eqref{2pfpossp}. Notice that, due to the anomalous dimension, we are led to  an ``Euclidean'' interpretation of the coordinate transformation. This interpretation should be contrasted with the more intrinsically ``Lorentzian'' formulae found e.g. in \cite{Dubovsky:2017cnj}, in analogy to which one would  interpret the integral over $J_+$ in \eqref{fdepcootr} as the total charge to the past of the line of constant  $x^+$, leading to $\Theta$-function shifts of the right-moving coordinate and no anomalous dimension\footnote{It is intriguing to note that if we plug into \eqref{transfcorr} the Lorentzian coordinate transformation that shifts $\bar z=x^-$ by the charge to the past of $z=x^+$, where $x^+$ is viewed as (null) time, we obtain a structure that is identical to the dipole star product \cite{Dasgupta:2001zu}.  }. 

%
%
%

When the anomaly term is taken into account $(k\neq 0)$, there is no nice interpretation of the two-point function in terms of a classical field\footnote{ The  Fourier transform of \eqref{2pftest} for $k\neq 0$ is a ${}_1F_1$ hypergeometric function whose argument is   $-\frac{2\pi^2(\bar z +i \mu q/\pi \ln z)^2}{\mu^2 k \ln z}$.}.
It is however easy to notice that if we replace the operators on the right-hand side of \eqref{transfcorr} by a  formal expansion where 
\be
 \O\left(z,\bar z -\frac{i \mu}{2\pi} \int^z\!\! dz'\, J_z(z') \right) \equiv  \tilde \O(z,\bar z) =\O (z,\bar z) - \frac{ i\mu}{2\pi} \int^z \! \! dz' :\! J(z') \, \bar \p \O(z,\bar z)\!: - \frac{\mu^2}{8\pi^2} \int \!:\! J \!\!\int\! J \, \bar \p^2 \O\!: + \ldots 
\ee
then compute the formal correlator
\bea
\langle \tilde \O(z_1,\bar z_1) \tilde \O (z_2, \bar z_2) \rangle_{CFT} &= & \left(1 + \frac{i\mu q}{2\pi}  \ln z_{12} (\bar \p_1 -  \bar \p_2) - \frac{\mu^2 k}{8\pi^2} \ln z_{12} \bar \p_1 \bar \p_2 - \frac{\mu^2 q^2}{8\pi^2} \ln^2 z_{12} (\bar \p_1 -  \bar \p_2)^2 + \ldots  \right) \nonumber \\  
& \times& \langle  \O(z_1,\bar z_1)  \O (z_2, \bar z_2) \rangle_{CFT} 
\eea
and   Fourier transform, it will reproduce the normalized two-point function   \eqref{2pftest}. We conclude that the coordinate transformation we need to reproduce the full two-point function involves an integral over the \emph{operator} $J$, and that the shift by this operator-valued quantity needs to be carefully defined. 
%
 It would be very interesting to ascertain whether similar subtleties play a role in the  $T\bar T$ deformation. 
 

The next step is to check whether the three-point functions also obey a relation of this sort. It is easy to check that the dimension shifts in \eqref{3pf}, 
including the contribution of the anomaly, will be  reproduced by the formal correlator 

\be
\langle \tilde \O(z_1,\bar z_1) \tilde \O (z_2, \bar z_2)\tilde \O (z_3, \bar z_3) \rangle_{CFT} 
\ee
provided 
the correction to the OPE coefficients \eqref{corrOPE}  vanishes to all 
orders in  conformal perturbation theory. If this turns out to be the case, then  the $J\bar T$ deformation can also be interpreted as a  spectral flow transformation,  where the flow parameter depends on $\bar p$. 
%
However, as we will show in section \ref{lincorr3pf}, the corrections to the OPE coefficients do not vanish, at least for the (standard) way in which we have chosen to deal with the UV divergences. Thus, our current results do not support an interpretation of general correlators in $J\bar T$-deformed CFTs in terms of operator-dependent coordinate transformations.



\section{The spectrum of $J\bar T$ deformed CFTs on the plane \label{specboost}}

In this section, we derive  the spectrum of  conformal dimensions of $J\bar T$ deformed CFTs on the plane, starting from the known spectrum of energies on the cylinder \cite{Guica:2017lia,kutasov,cssint}. 

In a two-dimensional CFT, the two spectra are simply related by the exponential map from the cylinder to the plane, $z_{pl}=\exp(2\pi i z_{cyl}/R)$ and its complex conjugate, which are conformal transformations and thus  symmetries of the theory. 
For a $J\bar T$-deformed CFT, only the holomorphic part of the   exponential map is  a symmetry of the action, as the latter is only invariant under $SL(2,\mathbb{R})_L \times U(1)_R$. This symmetry is  broken  to  $U(1)_L \times U(1)_R$ when placing the theory on a cylinder, due to the simultaneous action of the cylinder identifications on $z_{cyl}$ and $\bar z_{cyl}$. 
%
 The idea of this section is to first find the spectrum of  $J\bar T$ deformed CFTs on an infinitely boosted cylinder, on which the $SL(2,\mathbb{R})_L $ symmetry is restored,
%
  and then use the holomorphic exponential map to recover the conformal dimensions.

We start this section by briefly reviewing the spectrum of  $J\bar T$ deformed CFTs on the cylinder, and then perform an infinite boost to obtain a prediction for the exact spectrum on the plane.   

\subsection{Review of the spectrum on the cylinder}

We consider a $J\bar T$-deformed CFT on a cylinder of circumference $R$, i.e. with the spatial coordinate identified as $\varphi \sim \varphi +R$. 
The change in energy levels and chiral $U(1)$ charge as the deformation parameter is varied is 

\be
\frac{\p E}{\p \mu} = 2 R \, \langle J_z \rangle \langle T_{\bar z \bar z} \rangle = - Q \left( \frac{\p E}{\p R} + \frac{P}{R}\right)
\ee

\be
 \frac{\p Q}{\p \mu} = \frac{k}{4\pi} R\, \langle T_{\bar z \bar z} \rangle = -  \frac{k}{4\pi} R \, \p_R E_R
\ee
where $P$ is the total momentum (quantized in units of $1/R$), $E_R= \frac{1}{2} (E-P)$ is the right-moving energy, and we are using the conventions of \cite{cssint}. Using the fact that  the only dimensionless parameter in the problem is $\mu R$, the above equations can be written as

\be
\p_\mu E_R = - Q \,\p_R E_R \;, \;\;\;\;\;\; \p_\mu Q = 
\frac{k}{4\pi} (E_R + \mu \, \p_\mu E_R)
\ee
and imply that 
\be
E_{R} - \frac{2\pi Q^2}{k R} = const. \;, \;\;\;\;\;\;
Q= Q_0 + \frac{\mu k}{4\pi} E_R \label{enchshift}
\ee
Plugging in the expression for $Q$, the first equation becomes 

\be
 E_R - \frac{2\pi}{k R} \left( Q_0 + \frac{\mu k}{4\pi} E_R\right)^2 = \frac{2\pi}{R} \left( \bar h - \frac{c}{24}- \frac{Q_0^2}{k} \right)
\ee
where $\bar h$ denotes the right-moving conformal dimension associated to the corresponding state in the undeformed CFT.  The solution for the right-moving energy in terms of the original CFT data is

\be
E_R = \frac{4\pi}{\mu^2 k} \left( R - \mu Q_0 - \sqrt{\left(R-\mu Q_0 \right)^2 - \mu^2 k \left(\bar h- \frac{c}{24} \right)}\right) \label{er}
\ee
The expression for $E_L$  follows from momentum conservation

\be
E_L = E_R + \frac{2\pi}{R} ( h - \bar h) \label{el}
\ee

\subsection{The spectrum on  an infinitely boosted cylinder }

Let $x^\pm = \varphi \pm t$ be the coordinates on the cylinder, which above are identified as $x^\pm \sim x^\pm + R$. We would like to derive the spectrum on a boosted cylinder, with the identifications

\be
\tilde x^+ \sim \tilde x^+ + \tilde R_+ \;, \;\;\;\;\;\; \tilde x^- \sim \tilde x^- + \tilde R_-
\ee
where $\tilde R_+$ is finite and $\tilde R_- \r \infty$. The spectrum of $J\bar T$ deformed CFTs on a space with these identifications can be related to the spectrum on the usual cylinder via a boost

\be
\tilde x^\pm \r x^\pm = e^{\pm \g}  \tilde x^\pm
\ee
with the boost parameter chosen such that 

\be
  \tilde R_+ \, e^\g = \tilde R_-\, e^{-\g} =R \label{boostedrad}
\ee
 The various quantities before and after the boost are related as

\be
\mu = \tilde \mu \, e^{-\g} \;, \;\;\;\;\; E_L = e^{-\g} \tilde E_L \;, \;\;\;\;\;\; E_R = e^{\g} \tilde E_R \label{repl}
\ee
where the transformation law for $\mu$ follows from the fact that it is a constant null vector with a lower `$+$' component. $E_{L,R}$ are the energies on the unboosted cylinder, given by 
\eqref{er} and \eqref{el}. 

We will be interested in the limit $\g \r \infty$ with $\tilde \mu$, $\tilde R_+$ and the right-moving energy/momentum  $\tilde E_R$ fixed. The expression for $\tilde E_L$ in terms of these fixed  quantities is


\be
\tilde E_L = e^\g \left(E_R + \frac{2\pi ( h - \bar h)}{R} \right) =\frac{2\pi \left( h -c/24\right) }{\tilde R_+} + e^{2\g} \left(\tilde E_R- \frac{2 \pi (\bar h -c/24)\, e^{-2\g}}{\tilde R_+}\right)
\ee
The only way that $\tilde E_L$ could be finite is if $\tilde E_R$ cancels against $2\pi (\bar h -c/24)\, e^{-2\g}/\tilde R_+$ with precision $e^{-2\g}$. Since $\tilde E_R$ itself is finite, we conclude that we should scale

\be
\bar h - \frac{c}{24} = \bar h_0 \, e^{2\g} \label{ho}
\ee
with $\bar h_0$ fixed as $\g \r \infty$. The expansion of $\tilde E_R$ in this limit, using the replacements \eqref{boostedrad}, \eqref{repl} and \eqref{ho}, is given by
\bea
\tilde E_R &= & \frac{4\pi e^{-\g}}{\tilde \mu^2 k e^{-2\g}} \left(\tilde R_+ e^\g - \tilde \mu \,  Q_0 e^{-\g} - \sqrt{(\tilde R_+ e^\g - \tilde \mu \, Q_0 e^{-\g})^2 - \tilde \mu^2 k \bar h_0}\right) \nonumber \\
&=& \frac{2 \pi \bar h_0}{\tilde R_+} + \frac{2\pi e^{-2 \g}}{\tilde R_+} \left( \tilde \mu \, Q_0 \frac{\bar h_0}{\tilde R_+}+ \frac{\tilde \mu^2 k}{4} \left(\frac{\bar h_0}{\tilde R_+}\right)^2\right) + \mathcal{O}(e^{-4 \g})
\eea
We see that $\tilde E_R$ equals indeed the finite quantity $\frac{2 \pi \bar h_0}{\tilde R_+}$ within $e^{-2\g}$ precision. The  $\O(e^{-4\g})$ term will not survive the $\g \r \infty$ limit inside $\tilde E_L$, so we can neglect it. Defining the right-moving momentum

\be
\bar p \equiv \tilde E_R =   \frac{2 \pi \bar h_0}{\tilde R_+}
\ee
we find that
\be
\tilde E_L = \frac{2\pi ( h -c/24)}{\tilde R_+} + \frac{2\pi}{\tilde R_+} \left(\tilde \mu\, Q_0 \frac{\bar p}{2\pi} + \frac{\tilde \mu^2 k}{4 \cdot 4 \pi^2} \bar p^2\right)
\ee
Using the usual map from the cylinder  to the  plane, which maps energies to conformal dimensions, the anomalous left dimensions we find are

\be
h(\mu) = h + \frac{\mu}{2\pi} q \bar p + \frac{\mu^2 k}{16 \pi^2} \bar p^2
\ee
where we have dropped the tilde from $\mu$ and replaced the initial charge $Q_0$ by $q$.  The expression for the charge in the deformed theory is given by \eqref{enchshift}, which in our new notation  reads

\be
q(\mu) = q + \frac{\mu k}{4 \pi} \bar p \label{chshift}
\ee
The above expressions correspond to an exact formula for the conformal dimensions and charges in $J\bar T$-deformed CFTs as a function of the deformation parameter. As expected, these data only depend  on the combination $\mu \bar p$ and, unlike the spectrum on the cylinder, show no obvious pathology as $\mu$ becomes large.    Remarkably, these exact expressions terminate at $\O(\mu^2)$. The combination $\hat h \equiv h(\mu) - q(\mu)^2/k$
%
%
is independent of $\mu$, in perfect agreement with the argument presented in \cite{kutasov}, 
and hints towards a possible interpretation of the $J\bar T$ deformation as  spectral flow.

A simple consequence of the above formulae is that
purely holomorphic quantities do not acquire an anomalous dimension, as they have $\bar p =0$; however, initially purely antiholomorphic quantities, such as the right-moving stress tensor,  receives a non-trivial left-moving anomalous dimension and charge

\be
h_{\bar T}(\mu)= \frac{\mu^2 k \bar p^2}{16\pi^2}\;, \;\;\;\;\; q_{\bar T}(\mu) = \frac{\mu k \bar p}{4\pi} \label{hqtbar}
\ee
In the following section, we will check these formulae to leading order in conformal perturbation theory, finding perfect agreement.

\section{Conformal perturbation theory calculations \label{CPT}}

In this section, we present a selection of explicit conformal perturbation theory calculations, which follow the outline of section \ref{cptsetup}. First, we compute the leading order correction to the dimension and charge of an operator and show that it is in perfect agreement with the result of the previous section.  Next, we compute the linearized correction to a set of three-point functions. 
 Finally, we discuss the issue of operator mixing and perform several checks of the general structure discussed in section \ref{cptsetup}.
 

\subsection{Leading order correction to the dimension and the charge}


 To find the leading order shift in the conformal dimension and the charge, we need to compute the correlators $
 \langle \O\O^\dag\rangle$ and $
 \langle J \O \O^\dag \rangle$
to linearized or second order in $\mu_E$. The anomalous dimension can be read off from the coefficient of the  $\ln z$ term in the two-point function, while the charge is read off from the correlator with the current. The deforming operator we use is just the $J\bar T$ operator of the original CFT, since the corrections to it only become relevant at cubic order and higher. 



\subsubsection*{Anomalous dimension of a generic operator at linear order}

To first order in $\mu_E$, the correction to the two-point function of a generic operator is\footnote{We will oftentimes  replace the $(z_i,\bar z_i)$ arguments by a subscript. 
The normal ordering symbol around $J\bar T$ is not truly needed, as the two operators do not have a singular OPE, but we find this notation useful  for emphasizing the insertions of the deforming operator. 
}

\be
\d_1 \langle \O (z_1,\bar z_1) \O^\dag (z_2, \bar z_2)\rangle = \mu_E \int d^2 z_3\, \langle \O_1 \O_2^\dag    :\!J \bar T_3\!: \rangle = \frac{\mu_E q \bar h}{z_{12}^{2h-1}  \bar z_{12}^{2 \bar h-2} } \int \frac{dz_3 d\bar z_3}{z_{23} z_{13}\bar z_{23}^2 \bar z_{13}^2 } \label{intanomd}
\ee
The integral in \eqref{intanomd} can be written as $\p_{\bar z_1} \p_{\bar z_2} \mathcal{I}$ of the basic integral

\be
\mathcal{I}(z_1,z_2) = \int   \frac{d^2 z_3}{|z_{13}|^2 |z_{23}|^2}  = \frac{4\pi}{|z_{12}|^2} \left( \frac{2}{\e} + \ln |z_{12}|^2 + \g + \ln \pi  + \O(\e)\right) \label{iint}
\ee
which is computed in appendix \ref{useint}, equation \eqref{iint}, using 
dimensional regularization with $d=2 + \e$.  The first order change in the correlator thus reads

\be
\d_1 \langle \O (z_1, \bar z_1) \O^\dag  (z_2, \bar z_2)\rangle = - \frac{8 \pi \mu_E q \bar h}{z_{12}^{2h} \bar{z}_{12}^{2\hbar +1}} \left( \frac{2}{\e} + \ln |z_{12}|^2 + \g + \ln \pi  -\frac{3}{2}+ \O(\e)\right) \label{1stordcorroo}
\ee
It is now useful to write the operators in terms of the $(z,\bar p)$ variables by first analytically continuing $z,\bar z$ to real values and then
 performing the Fourier transform with respect to $\bar z_1, \bar z_2$. We find   
\bea
\d_1 \langle \O_{\bar p} (z_1) \O_{-\bar p}^\dag  (z_2)\rangle &=&  4 \pi iq  \mu_E  \bar p \left( \frac{2}{\e} + \ln z_{12} + \g + \ln \pi  -\frac{3}{2} + \psi (2 \bar h+1) - \ln |\bar p| + \frac{i\pi}{2} + \O(\e)\right) \times \nonumber \\
&& \hspace{0.6cm}\times \; \langle  \O_{\bar p} (z_1) \O_{-\bar p}^\dag  (z_2)\rangle_0 
\eea
where we used the Fourier transforms listed in appendix \ref{fourier}.  
Next, we define the renormalized operator

\be
\O^{ren}_{\bar p}(z) = \O_{\bar p}(z) + i \mu_E \bar p \, (c_\e^\O L^{\e})  \, \O_{\bar p} (z)  \label{oren}
\ee
where $L$ is a length scale that compensates the shift by $\e$ in the dimension of $\mu_E$.
If $c_\e^\O$ is constant, this simply corresponds to a redefinition  $\O_{ren} = \O +\mu_E\, c_\e L^{\e}  \p_{\bar z} \O  $ in position space. 
The coefficient $c_\e^\O$ is chosen to absorb the $2/\e$ divergence, as well as any finite terms needed to yield  some particular choice of normalization for the two-point function. Choosing the normalization \eqref{2pfnorm}, the associated coefficient is given by

\be
 c_\e^{\O} =  - 2 \pi q \left( \frac{2}{\e} + \g + \ln \pi  -\frac{3}{2}+ \psi (2 \bar h+1) - \ln |\bar p| L + \frac{i\pi}{2} \right) = - c_\e^{\O^\dag} 
\ee
Absorbing the $\ln \bar p$ term in the normalization is rather unusual from the point of view of position space interpretation of the operator. However, since here $\bar p$ is simply a label for the operator and the theory is non-local on the right, one may argue that such a redefinition should be allowed.  
 Using the relation $\mu_E =  i \mu/(4\pi^2)$ between the Euclidean and Lorentzian deformation parameters, the  two-point function of the renormalized operators reads

%
%

\be
\d_1 \langle \O^{ren}_{\bar p}(z_1) \O^{ren}_{-\bar p}(z_2) \rangle =  \left(1- \frac{  q \mu \bar p}{\pi}\, \ln  \frac{z_{12}}{L}\right) \langle \O_{\bar p}(z_1) \O_{-\bar p}(z_2) \rangle_{CFT}
\ee
The term in paranthesis corresponds to an anomalous dimension of $\d_1 h = q \mu \bar p/(2\pi) $. 
This perfectly matches the prediction of section \ref{specboost} for the linear correction to the  dimension of a charged operator.

\subsubsection*{Charge shift of a generic operator at linear order}

We would now like to reproduce the linear correction \eqref{chshift} to the charge of an operator from a conformal perturbation theory calculation. 
For this, we need to evaluate the three-point function $\langle J \O \O^\dag \rangle$ at order $\mu$. Using the fact that the current is unchanged to leading order in perturbation theory, the first order change in this three-point function should take the following form
\be
\d_1 \langle  \O_1 \O_2^\dag J_3 \rangle =  \left(\frac{q}{z_{31}} - \frac{ q}{z_{32}} \right)\d_1 \langle \O_1 \O_2^\dag \rangle + \left(\frac{\d_1 q_\O}{z_{31}} + \frac{\d_1 q_{\O^\dag}}{z_{32}} \right)\langle \O_1 \O_2^\dag \rangle_{CFT} \label{genfjoo}
\ee
where $\d_1 \langle \O_1 \O_2^\dag\rangle$ was computed in \eqref{1stordcorroo} and $\d_1 q_\O$ and $\d_1 q_{\O^\dag}$ represent the first-order correction to the charge of $\O$ and respectively $\O^\dag$.

The linearized correction to the three-point function is given by
\be
\d_1 \langle \O_1 \O^\dag_2 J_3  \rangle =  \mu_E\! \int \! d^2 z_4 \, \langle \O_1 \O^\dag_2 J_3 :\! J \bar T_4\!: \rangle  = \, \frac{\mu_E  \bar h}{z_{12}^{2h} \bar{z}_{12}^{2\hbar-2} }\left(\frac{k}{2} \int \frac{d^2 z_4}{z_{34}^2 \bar{z}_{14}^2 \bar z_{24}^2} + \frac{q^2 z_{12}^2}{z_{13} z_{23}} \int\frac{d^2 z_4}{z_{14} z_{24} \bar{z}_{14}^2 \bar z_{24}^2} \right) \label{jooint}
\ee
The first integral can be performed by taking sums and derivatives of the basic integral
\be
I_{ij} = \int \frac{d^2 z_4}{z_{4i} \bar z_{4j}}= - 2 \pi \left( \frac{2}{\e}+ \ln |z_{12}|^2 + \g_E + \ln \pi + \O(\e)\right) \equiv - 2 \pi (\ln |z_{12}|^2 + c_\e) \label{iij}
\ee
and is calculated in the appendix \ref{useint}, equation \eqref{joointev} as an example.  The second integral is $\p_{\bar z_1} \p_{\bar z_2} \mathcal{I}(z_1,z_2)$, as before. Thus, the correlator evaluates to

\be
\d_1  \langle \O_1 \O^\dag_2 J_3  \rangle  =  \frac{4 \pi\mu_E  \bar h}{z_{12}^{2h} \bar{z}_{12}^{2\hbar+1}} \left[ \frac{ k}{2} \left(\frac{1}{z_{31}}- \frac{1}{z_{32}} \right) - 2q^2 \left(\frac{1}{z_{31}}- \frac{1}{z_{32}} \right)   \left( \frac{2}{\e} + \ln |z_{12}|^2 + \g + \ln \pi  -\frac{3}{2}\right)\right]
\ee
Next, we Fourier transform with respect to $\bar z_{1,2}$. Since the current is purely holomorphic, $\bar p_3 =0$, so by the momentum conservation equation $\O$ and $\O^\dag$  have equal and opposite momenta, $\bar p$ and $-\bar p$. After introducing the renormalized operators \eqref{oren} and trading $\mu_E$ for $\mu$, we find

\be
\d_1 \langle \O^{ren}_{\bar p}(z_1) \O^{ren\, \dag}_{-\bar p}(z_2) J(z_3)  \rangle =  \left(\frac{1}{z_{31}}- \frac{1}{z_{32}} \right)\, \left[ q \, \d_1 \langle \O^{ren}_{\bar p}(z_1) \O^{ren\,\dag}_{-\bar p}(z_2) \rangle + \frac{ \mu k  \bar p}{4\pi}\, \langle \O_{\bar p}(z_1) \O_{-\bar p}^\dag (z_2) \rangle_{CFT}\right]
\ee
This correlator is precisely of the general form \eqref{genfjoo}, where the second term corresponds to an $\O(\mu)$ shift in the charge

\be
\d_1 q_\O = \frac{k \mu \bar p}{4\pi}
\ee
which is equal and opposite for $O$ and $\O^\dag$.   Notice that holomorphy of  $J$ is
essential for maintaining charge conservation.

\subsubsection*{Second order charge shift of a generic operator}

The change in the $\langle J \O \O^\dag \rangle$ correlator to second order in the deformation is given by

\be
\d_2 \langle J_3 \O_1 \O_2^\dag \rangle = \frac{\mu_E^2}{2}   \langle J_3 \O_1 \O^\dag_2 \! \int\! d^2 z_4 :\! J\bar T_4\!: \! \int \! d^2 z_5 :\! J\bar T_5 \!:\rangle
\ee
We can use the OPE of $J_3$ with the various other insertions to rewrite this correlator as

\be
\d_2 \langle J_3 \O_1 \O_2^\dag \rangle = q \left(\frac{1}{z_{31}} - \frac{1}{z_{32}} \right) \d_2 \langle  \O_1 \O_2^\dag \rangle +\frac{ k \mu_E}{2} \int \frac{d^2 z_4}{z_{34}^2}\, \d_1 \langle  \O_1 \O_2^\dag \,\bar T_4 \rangle
\ee
%
%
%
%
%
The last term can be evaluated by using the OPE  for $\bar T_4$
\bea
 &&\hspace{-1 cm}\int \frac{d^2 z_4}{z_{34}^2}\, \d_1 \langle  \O_1 \O^\dag_2 \,\bar T_4 \rangle = \int \frac{d^2 z_4}{z_{34}^2} \left(\frac{\bar h}{\bar z_{41}^2} + \frac{1}{\bar z_{41}} \p_{\bar z_1} + \frac{\bar h}{\bar z_{42}^2} + \frac{1}{\bar z_{42}} \p_{\bar z_2}  \right)\d_1 \langle  \O_1 \O_2^\dag  \rangle + \\
 && \hspace{1 cm} + \mu_E \int \frac{d^2 z_4 d^2 z_5}{z_{34}^2} \left[ \frac{c}{2 \bar z_{45}^4} \langle \O_1 \O_2^\dag J_5 \rangle_{CFT} + \left(\frac{2}{\bar z_{45}^2} + \frac{1}{\bar z_{45}} \p_{\bar z_5}  \right)\langle \O_1 \O_2^\dag J_5 \bar T_5 \rangle_{CFT} \right]\nonumber \\
 &&\hspace{-0.9 cm} = 
 \frac{2 \pi}{z_{13}} \d_1 \langle \bar \p \O_1 \O_2^\dag \rangle + \frac{2 \pi}{z_{23}} \d_1 \langle  \O_1 \bar \p \O_2^\dag  \rangle_{n-1}+
 \frac{\pi^2 \mu_E  c}{3} \p^2_{\bar z_3}\langle \O_1 \O_2^\dag J_3 \rangle_{CFT} + 4 \pi^2 \mu_E \langle \O_1 \O_2^\dag J_3 \bar T_3 \rangle_{CFT} \nonumber 
 \eea
where in the last line we have discarded  the contact terms proportional to $\bar h$. The end result is
\bea
\d_2 \langle J_3 \O_1 \O_2^\dag \rangle &=& q \left(\frac{1}{z_{31}} - \frac{1}{z_{32}} \right) \d_2 \langle  \O_1 \O_2^\dag \rangle - \frac{ \pi k \mu_E }{z_{31}} \d_1 \langle \bar \p \O_1 \O_2^\dag  \rangle - \frac{ \pi k\mu_E}{z_{32}} \d_1 \langle  \O_1 \bar \p \O_2^\dag  \rangle + \nonumber \\
&+&\frac{\pi^2 k\mu_E^2 c}{6} \p^2_{\bar z_3}\langle \O_1 \O_2^\dag J_3 \rangle_{CFT} + 2 \pi^2 k\mu_E^2 \langle \O_1 \O_2^\dag J_3 \bar T_3 \rangle_{CFT} 
\eea
Replacing $\bar \p_i \r i \bar p_i$, we immediately find the charge shifts we expect; in particular, the second order correction to the charge vanishes, in agreement with \eqref{chshift}. The first term on the second line is just a contact term, whereas the last term should be moved to the left-hand side and represents the quadratic correction the current $J^{corr} = J - 2 \pi^2 \mu_E^2 k :\! J\bar T\!:$ derived in \eqref{jredefmusq}. Notice that once we know the corrections to the deforming operator and to the chiral current at higher orders in $\mu$, the above calculation can be easily extended to higher orders to give a recursive formula for $\d_n \langle J \O \O^\dag \rangle$ in terms of the corrections at lower orders in conformal perturbation theory. 


\subsubsection*{Anomalous dimension of the right-moving stress tensor}

One of the predictions of our formula \eqref{defspec} for the spectrum is that the right-moving  stress tensor, which starts out as a purely antiholomorphic operator,  acquires a left-moving conformal dimension and charge given by \eqref{hqtbar}. In the following, we  reproduce this prediction to second order in $\mu$.


Since the linear correction to the $\langle \bar T \bar T \rangle$ two-point function vanishes, we need to 
evaluate

\be
\d_2 \langle \bar  T_1 \bar T_2 \rangle = \frac{\mu_E^2}{2} \int d^2 z_3 \, d^2 z_4 \langle \bar T_1 \bar T_2 :\! J \bar T_3\!: :\!J \bar T_4\! : \rangle
\ee
The integrand factorizes into a purely holomorphic part, which equals the current two-point function $k/(2 z_{34}^2)$, and a purely antiholomorphic part that corresponds to the correlation function of four stress tensors. The latter is given in appendix \ref{listcorr}, equation \eqref{stresst4pf}.  The integral then reads
\bea
\d_2 \langle \bar  T_1 \bar T_2 \rangle & = & \frac{\mu_E^2 k }{4} \int \frac{d^2 z_3 d^2 z_4}{z_{34}^2}\left[ \, c \left(\frac{1}{\bar z_{12}^2 \bar z_{13}^2 \bar z_{24}^2 \bar z_{34}^2} +  \frac{1}{\bar z_{12}^2 \bar z_{14}^2 \bar z_{23}^2 \bar z_{34}^2} + \frac{1}{\bar z_{13}^2 \bar z_{14}^2\bar  z_{24}^2\bar z_{23}^2}  \right) + \right. \nonumber \\
&& \hspace{1cm} \left. + \, \frac{c^2}{4} \left(\frac{1}{\bar z_{12}^4\bar z_{34}^4} + \frac{1}{\bar z_{13}^4 \bar z_{24}^4} + \frac{1}{\bar z_{23}^4\bar z_{14}^4} \right) \right] \label{del2tbarsq}
\eea
It is simplest to express the end result in terms of the basic integral \eqref{iij}

\be
\d_2 \langle \bar T_1 \bar T_2 \rangle = \frac{\mu_E^2 k }{4} \left(2 \, c \cdot \frac{12 \pi I_{12} + 44 \pi^2}{\bar z_{12}^6} + c \cdot \frac{16 \pi (I_{12} + 3 \pi)}{\bar z_{12}^6} + 2\, \frac{c^2}{4} \cdot \frac{40 \pi^2}{3 \bar z_{12}^6}\right) \label{tbar2pfmusq}
\ee
The provenance of each term from \eqref{del2tbarsq} should be clear. Passing to momentum space, we find

\be
\d_2 \langle \bar T_{\bar p} \,(z_1) \bar T_{-\bar p} \, (z_2) \rangle=   -\frac{ k \mu^2 \bar p^2}{8 \pi^2 }\,\left(\ln z_{12} +  c_\e - \frac{17}{10} - \frac{c}{12}+\psi (6) - \ln |\bar p| + \frac{i \pi}{2}\right)\,  \langle \bar T_{\bar p} \, (z_1) \bar T_{-\bar p}\, (z_2) \rangle_{CFT} \,
\ee
From the log term, we can read off the holomorphic anomalous dimension of the stress tensor 

\be
h_{\bar T} = \frac{k \mu^2 \bar p^2}{16 \pi^2}
\ee
which precisely agrees with our expectations.  Defining the renormalized operators
\be
\bar T^{ren}_{\bar p} (z) = \bar T_{\bar p} \, (z) - c_{\bar T} \mu^2 
\bar p^2 \bar T_{\bar p}
\ee
with $c_{\bar T} = - \frac{k}{16\pi^2} (c_\e -\frac{17}{10}-\frac{c}{12}+\psi(6)+ \frac{i\pi}{2} - \ln \bar p)$, the two-point function   can be put  in the form \eqref{2pfnorm}.

\subsubsection*{Charge shift of the right-moving stress tensor}

The charge shift of the antiholomorphic stress tensor
is  visible already at first order in the perturbation. The integral that we need to perform is
\be
\d_1\langle J_3 \bar T_1 \bar T_2\rangle = \mu_E \langle J_3 \bar T_1 \bar T_2 \int d^2 z_4 \,: \!J \bar T_4 \!: \rangle = \frac{\mu_E kc}{2\bar z_{12}^2} \int \frac{d^2 z_4}{z_{34}^2\bar z_{14}^2 \bar z_{24}^2} =  \frac{\mu_E kc}{2\bar z_{12}^2} \cdot \frac{4\pi}{\bar z_{12}^3} \left( \frac{1}{z_{31}} - \frac{1}{z_{32}} \right)
\ee
Passing to momentum space, we find

\be
\d_1\langle J (z_3)\, \bar T_{\bar p}\, (z_1) \bar T_{-\bar p} \, (z_2)\rangle = \frac{\mu k \bar p}{4\pi}\left( \frac{1}{z_{31}} - \frac{1}{z_{32}} \right) \langle \bar T_{\bar p}\, (z_1) \bar T_{-\bar p} \, (z_2)\rangle_{CFT}
\ee
and thus $\bar T$ acquires the expected charge of $\mu k \bar p/(4\pi)$. Each of the $\bar T$ insertions in the correlator above carries an 
 equal and opposite charge, due to the fact that $J$ has zero right-moving momentum. 
 
 The  $\O(\mu^2)$ contribution to the $\langle J \bar T \bar T \rangle$ correlator vanishes trivially, which is consistent with the fact that the charge does not receive any $\O(\mu^2)$ correction. Indeed,

\be
\d_2 \langle J_3 \bar T_1 \bar T_2 \rangle = 0=\frac{z_{12}}{z_{31}z_{32}} \left( q_{\bar T} \d_2 \langle \bar T_1 \bar T_2 \rangle + \d_1 q_{\bar T} \d_1 \langle \bar T_1 \bar T_2 \rangle + \d_2 q_{\bar T} \langle \bar T_1 \bar T_2 \rangle_{CFT} + 2\pi^2 \mu_E^2 k\langle :\!J \bar T_3 \!: \bar T_1 \bar T_2\rangle_{CFT} \right) 
\ee
The first term on the right-hand side vanishes because $\bar T$ is originally neutral, the second because the linearized correction to the $\langle \bar T \bar T \rangle$ two-point function is zero  and the fourth because this particular correlator vanishes in the original CFT. We thus deduce that $\d_2 q_{\bar T} =0$, in agreement with \eqref{hqtbar}.

\subsection{Linear correction to the three-point functions \label{lincorr3pf}}

In this subsection, we would like to i) check that the structure of the three-point function computed to first order in conformal perturbation theory agrees with the general form proposed in section \ref{genstr} and ii) check whether the correction to the OPE coefficients is non-zero at linear order in $\mu$. The computations are performed for two different  three-point functions:  $\langle \bar T \O \O^\dag \rangle$ and the correlator  of three generic charged operators. 


\subsubsection*{Linearized correction to  $\langle \bar T \O \O^\dag \rangle$ }

Given the 
general form of the three-point functions \eqref{3pf}, the answer that we expect to obtain 
is the expansion to linear order in $\mu$ of

\be
\langle \bar T_3 \O_1 \O_2^\dag \rangle_{ren} = \frac{  K (2 \bar h -2,2,2)\, C_{\bar T \O \O} (\mu \bar p_i)}{z_{13}^{\mu q (\bar p_1 + \bar p_2)/2\pi} z_{23}^{-\mu q (\bar p_1 + \bar p_2)/2\pi} z_{12}^{2h+\mu q (\bar p_1-\bar p_2)/2\pi}} \label{too3pf}
\ee
where we have ommitted the $\O(\mu^2)$ corrections to the dimensions, which will not contribute, and also dropped the momentum-conserving delta function. The OPE coefficient has the expansion

\be
C_{\bar T \O \O} (\mu \bar p_i)=\bar h + \O(\mu \bar p_i) \label{ctoo}
\ee
We would now like to reproduce this form from a conformal perturbation theory calculation. The linearized correction to the bare correlator is
\bea
\d_1 \langle \bar T_3 \O_1 \O_2^\dag \rangle &=& \mu_E
\langle \bar T_3 \O_1 \O_2^\dag \int d^2 z_4 :\! J \bar T_4 \!: \rangle \nonumber \\
&=& \frac{\mu_E q}{z_{12}^{2h-1} \bar z_{12}^{2\bar h}} \left[\, \frac{\bar h^2 \bar z_{12}^4}{\bar z_{13}^2 \bar z_{23}^2 } \int \frac{d^2 z_4}{z_{14}z_{24} \bar z_{14}^2 \bar z_{24}^2} + \frac{2\bar h \bar z_{12}^2}{\bar z_{13} \bar z_{23} } \int \frac{d^2 z_4}{|z_{14}|^2 |z_{24}|^2 \bar z_{34}^2} + \frac{c}{2}\int \frac{d^2 z_4}{z_{14} z_{24} \bar z_{34}^4}\right] \nonumber \\
&=& \frac{\mu_E q}{z_{12}^{2h-1} \bar z_{12}^{2\bar h}} \left[\, \frac{2\bar h \bar z_{12}}{\bar z_{13}\bar z_{23} z_{12}}\left(\frac{1}{\bar z_{13}^2} (
I_{23}-I_{21}-I_{13})+ \frac{1}{\bar z_{23}^2} (
I_{13}-I_{12}-I_{23}) - \frac{2\pi \bar z_{12}^2}{\bar z_{13}^2\bar z_{23}^2}\right) + \right. \nonumber \\
&& \hspace{5 mm}+ \left. \frac{\bar h^2 \bar z_{12}^4}{\bar z_{13}^2 \bar z_{23}^2 } \p_{\bar z_1} \p_{\bar z_2} \mathcal{I}(z_{1,2}) + \frac{c \pi}{3 z_{12}} \left(\frac{1}{\bar z_{13}^3} - \frac{1}{\bar z_{23}^3}\right)\right] \label{del1tbaroo}
\eea
where $I_{ij}$ is as usual given by \eqref{iij}. 
This result  can be split into three parts:  one part proportional to $\ln z_{ij}$,  another part, with an  equal coefficient, multiplying $\ln \bar z_{ij}$ and a third part that   only involves powers of $z_{ij}$. After performing the Fourier transform with respect to the $\bar z_i$, the first part should give the shifts in the powers of $z_{ij}$ in \eqref{too3pf}, while the remaining two parts will combine, together with the renormalization of the operators, to give 
 the linearized correction to the OPE coefficient \eqref{ctoo}. 

\subsubsection*{\emph{Holomorphic logarithmic terms}}

To check that the coefficients of the $\ln z_{ij}$ terms agree with 
\eqref{too3pf}, we  need to collect the full coefficient of each logarithmic term, Fourier transform it with respect to the $\bar z_i$,  and then divide by the zeroth order contribution to the right-moving three-point function, which is $\bar h K(2\bar h -2,2,2)$.  We find: 
\bea
 \ln z_{13}:\; - \frac{4\pi \mu_E q \bar h  (\bar z_{13}+\bar z_{23})}{z_{12}^{2h} \bar z_{12}^{2\bar h -2} \bar z_{13}^3 \bar z_{23}^3} \;\;& \mbox{\Large{ $ \stackrel{\F_t\;}{\r}$}}& \;\;- \frac{4\pi \mu_E q \bar h  }{z_{12}^{2h} } \, [K(2\bar h-2,2,3)+K(2\bar h-2,3,2)] \nonumber \\
 &=&  \frac{2\pi i \mu_E q }{z_{12}^{2h}} (\bar p_1 +\bar p_2)\times  \bar h K(2\bar h-2,2,2) \label{coefflnz13}
\eea
%
Plugging in $\mu_E =  i \mu/(4\pi^2)$ and using the explicit expression \eqref{Kabc} for $K$, we find that the correction to the exponent of $z_{13}$ in the denominator is

\be
 + \frac{\mu q}{2\pi} (\bar p_1 +\bar p_2)
\ee
which perfectly matches our expectation. The coefficient of $\ln z_{23}$ is just minus the above, so
%
the correction to the exponent of $z_{23}$ is

\be
 - \frac{\mu q}{2\pi} (\bar p_1 +\bar p_2)
\ee
which again matches what we expect. Finally, the coefficient of 
\be
\ln z_{12}: \; \frac{4\pi q \mu_E \bar h }{z_{12}^{2h} \bar z_{12}^{2\bar h-1} \bar z_{13}^3 \bar z_{23}^3} (\bar z_{13}^2+\bar z_{23}^2 - 2 \bar h \bar z_{13} \bar z_{23} ) \;\;\mbox{\Large{ $ \stackrel{\F_t\;}{\r}$}} \;\; \frac{2\pi i q \mu_E }{z_{12}^{2h} } (\bar p_1 - \bar p_2)\times  \bar h K(2\bar h-2,2,2) \label{coefflnz12}
\ee
leading to the following correction to the exponent of $z_{12}$

\be
2h \r 2h + \frac{\mu q }{2\pi} (\bar p_1 -\bar p_2)
\ee
again in perfect agreement with \eqref{too3pf}.

\subsubsection*{\emph{Correction to the OPE coefficients}}

To compute  the correction to the OPE coefficients, we need to work in terms of the renormalized operators \eqref{oren}

\be
\O^{ren}_{\bar p} (z) = \O_{\bar p}(z) - 2 \pi i q \mu_E  \bar p\left(d_\e - \ln \bar p \right)  \O_{\bar p} (z)
\ee
where we have defined the constant $d_\e=\frac{2}{\e} + \g + \ln \pi - \frac{3}{2}+ \ln L + \psi(2\bar h+1)  + \frac{i \pi}{2} $.
The leading correction to the three-point function of renormalized operators  takes the form
\bea
\d_1  \langle \bar{T}_{\bar{p}_3}(z_3)\, \O^{ren}_{\bar{p}_1} (z_1) \, \O^{ren, \dag}_{\bar{p}_2} (z_2)  \rangle &=& \d_1 \langle \bar T_3 \O_1 \O_2^\dag  \rangle  + \nonumber \\
&& \hspace{-3.4cm}+\; 2 \pi i q \mu_E \left( \bar p_1  \ln \bar p_1-  \bar p_2  \ln \bar p_2 + (\bar p_2 - \bar p_1) d_\e  \right) \langle \bar T_3 \O_1 \O_2^\dag  \rangle_{CFT} \label{tbarooren}
\eea
where  we have assumed there is no renormalization of $\bar T$ to this order. All correlators above should be written in the $(z,\bar p)$ representation.  
The linearized correction $\d_1\langle \bar T_3 \O_1 \O_2^\dag\rangle $ is given in \eqref{del1tbaroo}.  Its holomorphic and antiholomorphic logarithmic pieces can be read off from \eqref{coefflnz13} and \eqref{coefflnz12}, while the part only  involving power laws reads
\be
\frac{\mu_Eq}{z_{12}^{2h} \bar z_{12}^{2\bar h}} \left[\, \frac{12 \pi \bar h^2 \bar z_{12}}{\bar z_{13}^2 \bar z_{23}^2 }- \frac{4\pi \bar  h \bar z_{12}^3}{\bar z_{13}^3 \bar z_{23}^3}+ \frac{\pi c}{3}\left(\frac{1}{\bar z_{13}^3}-\frac{1}{\bar z_{23}^3}\right) + c_\e \cdot \frac{4\pi \bar h \bar z_{12}}{\bar z_{13}^3 \bar z_{23}^3}(\bar z_{12}^2 - 2 (\bar h-1)\bar z_{13} \bar z_{23})\right] \label{powerlawtoo}
\ee
It is not hard to check that all the $1/\e$ divergences cancel in the renormalized correlator. 
Our task now is to compute the Fourier transform of the antiholomorphic log pieces and of \eqref{powerlawtoo} and  plug them into the renormalized correlator \eqref{tbarooren}. Note that the Fourier transforms of the logarithms will produce terms proportional to $\ln \bar p_i$. The resulting expression for the correction to the OPE coefficient is  not particularly illuminating, so we do not present it here;
%
it is however easy to see that it does not vanish, since:  i) the terms proportional to $c q$ in \eqref{powerlawtoo} have nothing against which to cancel and
ii) the $\ln \bar p_i$ terms do not cancel among each other, as we now show.
%
%
%

To compute the Fourier transform of $\ln \bar z_{ij}$, we  differentiate $-K(a,b,c)$ with respect to the parameters $a,b,c$. For example,
%
%
\be
 - \frac{4\pi \mu_E q \bar h  (\bar z_{13}+\bar z_{23})}{z_{12}^{2h} \bar z_{12}^{2\bar h -2} \bar z_{13}^3 \bar z_{23}^3} \ln \bar z_{13}\;\;\; \mbox{\Large{$\stackrel{\F_t\;}{\r}$}} \;\;\; \frac{4 \pi \mu_E q \bar h}{z_{12}^{2h}} [\p_c K (2\bar h -2, 3,2) + \p_c K (2\bar h -2,2,3)]
\ee 
Since in the above formula $b=2,3$, the hypergeometric function \eqref{Kabc} entering $K(a,b,c)$ has one negative argument, so  it reduces to a polynomial.
 The part proportional to $\ln \bar p_i$ is  given by

\be
- \frac{2 \pi i \mu_E q }{z_{12}^{2h}}  (\bar p_1 + \bar p_2) \ln \bar p_1 \, \bar h K(2\bar h-2,2,2) 
\ee
The Fourier transform of the term proportional to $\ln \bar z_{12}$ can be obtained in a similar fashion

\bea
\frac{4\pi q \mu_E \bar h }{z_{12}^{2h} \bar z_{12}^{2\bar h-1} \bar z_{13}^3 \bar z_{23}^3} (\bar z_{13}^2+\bar z_{23}^2 - 2 \bar h \bar z_{13} \bar z_{23} ) \ln \bar z_{12}
& \mbox{\Large{$\stackrel{\F_t\;}{\r}$}}   &  -\frac{2\pi i q \mu_E  }{z_{12}^{2h} } (\bar p_1-\bar p_2) \,\bar h K (2\bar h -2,2,2) \, \ln  \bar p_1  \nonumber 
\eea
Notice that the term proportional to $\bar p_2 \ln \bar p_1$ cancels between these two contributions. Even though in principle,  our procedure gives the exact answer for the Fourier transform,  the $\ln \bar p_i$ dependence can be simply derived from that of the $\bar p$-dependent prefactors that multiply the hypergeometric function  in \eqref{Kabc}.  
 
The Fourier transform of the term proportional to $\ln \bar z_{23}$ is

\be
\frac{4\pi \mu_E q \bar h  (\bar z_{13}+\bar z_{23})}{z_{12}^{2h} \bar z_{12}^{2\bar h -2} \bar z_{13}^3 \bar z_{23}^3}  \, \ln \bar z_{23}\;\;  \stackrel{\F_t}{\r} \;\;- \frac{4\pi \mu_E q \bar h  }{z_{12}^{2h} } [\p_b K(2\bar h -2, 3,2) + \p_b K (2\bar h -2, 2,3)]
\ee
is slightly more involved, since the derivative with respect to $b$ is taken before setting $b=2,3$. However, in appendix \ref{fourier} we show that the derivative  of the hypergeometric with respect to $b$  does not produce a $\ln \bar p$ term. Thus, the only logarithmic terms come from the $\bar p_2^{b-1}$ prefactor in \eqref{Kabc}, and we obtain

\be
\frac{2\pi i \mu_E q}{z_{12}^{2h}} (\bar p_1 + \bar p_2) \ln \bar p_2\, \bar h K (2\bar h -2,2,2)
\ee
Summing up the $\ln \bar p_i$ contributions in \eqref{tbarooren}, we find

\bea
\d_1  \langle \bar{T}_{\bar{p}_3}(z_3) \O^{ren}_{\bar{p}_1} (z_1) \O^{ren, \dag}_{\bar{p}_2} (z_2)  \rangle_{\ln \bar p} &=&
\frac{2\pi i \mu_E q }{z_{12}^{2h}} \left[(\bar p_1 + \bar p_2) \ln \bar p_2 - 2 \, \bar p_1 \ln \bar p_1+(  \bar p_1  \ln \bar p_1-  \bar p_2  \ln \bar p_2) \right] \times \nonumber \\
\times \; \bar h K (2\bar h -2,2,2) 
&=& - 2\pi i \mu_E q  \, \bar p_1 \ln \frac{\bar p_1}{\bar p_2}\,  \langle  \bar{T}_{\bar{p}_3}(z_3) \O_{\bar{p}_1} (z_1) \O^{\dag}_{\bar{p}_2} (z_2)  \rangle_{CFT} \label{logtoo}
\eea
Thus, we find that the correction to the OPE coefficient is non-zero for structural reasons, which follow from the particular way in which we have regulated the UV divergences of the integrated correlator. In addition, there are power law terms that show no particular structure, including the ones proportional to $cq$ that we mentioned earlier.

\subsubsection*{Linearized corrections to a generic three-point function}

 
Let us now compute the linearized corrections to a correlation function of three arbitrary charged operators, with charges $q_{1,2,3}$ (which must sum to zero) and right-moving dimensions $\bar h_{1,2,3}$. The correlator that we need to integrate is
\be
\langle \O_1 \O_2 \O_3 :\! J \bar T_4\!: \rangle = \left( \frac{q_1}{z_{41}}+ \frac{q_2}{z_{42}}+\frac{q_3}{z_{43}}\right)  \left(\frac{\bar h_1 \bar z_{12}\bar z_{13}}{\bar z_{14}^2 \bar z_{24} \bar z_{34}}-\frac{\bar h_2 \bar z_{12}\bar z_{23}}{\bar z_{14} \bar z_{24}^2 \bar z_{34}}+\frac{\bar h_3 \bar z_{13}\bar z_{23}}{\bar z_{14} \bar z_{24} \bar z_{34}^2} \right) \langle \O_1 \O_2 \O_3 \rangle_{CFT}
\ee
Note that it is sufficient to compute the terms proportional to $q_1$, the rest follow by permutation symmetry. This contribution is given by

\be
-\mu_E q_1 \left( \bar h_1 \bar z_{12} \bar z_{13} \p_{\bar z_1} -\bar h_2 \bar z_{12} \bar z_{23} \p_{\bar z_2}+\bar h_3 \bar z_{13} \bar z_{23} \p_{\bar z_3}\right) \int \frac{d^2 z_4}{z_{41} \bar z_{41} \bar z_{42} \bar z_{43}}
\ee 
where the integral evaluates to

\be
\int \frac{d^2 z_4}{z_{41} \bar z_{41} \bar z_{42} \bar z_{43}} 
= \frac{1}{\bar z_{23}} \left(\frac{I_{13}}{\bar z_{13}} - \frac{I_{12}}{\bar z_{12}} \right)
\ee
The full contribution proportional to $q_1$ then is
\bea
&& 2\pi q_1 \mu_E \left[ \frac{\bar z_{13} (\bar h_3 - \bar h_1) + \bar h_2 (\bar z_{12}-\bar z_{23})}{\bar z_{12} \bar z_{23}} (\ln |z_{12}|^2 + c_\e) + \frac{(\bar h_1-\bar h_2) \bar z_{12}- \bar h_3 (\bar z_{13}+\bar z_{23}) }{\bar z_{13} \bar z_{23}} (\ln |z_{13}|^2 + c_\e) \,+\right. \nonumber \\
&& \hspace{2.5cm} \left. + \, \frac{\bar h_3 \bar z_{12} + \bar h_2 \bar z_{13} + \bar h_1 (\bar z_{12} + \bar z_{23})}{\bar z_{12} \bar z_{13}} \right] \langle \O_1 \O_2 \O_3 \rangle_{CFT}
\eea
As already mentioned, the contributions proportional to $q_2$ and $q_3$ are obtained via permutations. Let us now check that the coefficients of the $\ln z_{ij}$ terms are consistent with the shifts in the anomalous dimensions. Explicitly, the coefficient of 
\bea
\ln z_{13} \;\;: \!\!\!&& 2 \pi q_1 \mu_E  \left( (\bar h_1 - \bar h_2) K (a-1,b+1,c+1) - \bar h_3 K(a,b,c+1) - \bar h_3 K(a,b+1,c) \right) \, + \nonumber \\
&+&  2 \pi q_3 \mu_E  \left( (\bar h_2 - \bar h_3) K (a+1,b-1,c+1) + \bar h_1 K(a+1,b,c) + \bar h_1 K(a,b,c+1) \right) \nonumber \\
&=& 2 \pi i \mu_E [q_1 (\bar p_1 + \bar p_2) - q_3 \bar p_1] \, K(a,b,c)
\eea
where $a,b,c$ are given by \eqref{abc} and the simplification is a consequence of hypergeometric identities. Thus, the correction to the exponent of $z_{13}$ takes the form

\be
\d_1 h_{13;2} = \frac{\mu}{2\pi}(q_1 (\bar p_1 + \bar p_2) - q_3 \bar p_1)
\ee
This should be compared to the expected

\be
\d_1 (h_1 + h_3 - h_2) = \frac{\mu}{2\pi} (q_1 \bar p_1 + q_3 \bar p_3 - q_2 \bar p_2)
\ee
The two expressions are indeed equal after using the charge conservation equation to write $q_2 = -q_1-q_3$, and momentum conservation to eliminate $\bar p_3$.  Thus, the coefficient of the $\ln z_{13}$ term exactly agrees with the prediction based on one-dimensional conformal symmetry.  The terms proportional to $\ln z_{12}$ and $\ln z_{23}$ can also be shown to perfectly match the CFT$_1$ prediction.

 As far as the corrections to the OPE coefficients are concerned, we can easily show that they do not vanish, by again focussing on the terms proportional to $\ln \bar p_i$. Concretely, the linearized  three-point function of renormalized correlators is given by
 \bea
\d_1\langle \O^{ren}_{\bar p_1} \O^{ren}_{\bar p_2} \O^{ren}_{\bar p_3}\rangle & = & \d_1 \langle \O_{\bar p_1} \O_{\bar p_2} \O_{\bar p_3} \rangle   -\frac{\mu}{2\pi} \left(q_1 \bar p_1 \ln \bar p_1+ q_2 \bar p_2 \ln \bar p_2 + q_3 \bar p_3 \ln \bar p_3 -  \right. \nonumber \\ &-&  \left. d_{\varepsilon} (q_1 \bar p_1 + q_2 \bar p_2 + q_3 \bar p_3)\right)  
 \, \langle \O_{\bar p_1} \O_{\bar p_2} \O_{\bar p_3} \rangle_{CFT}
\eea
 Concentrating on the pieces proportional to $\ln \bar p_i$, we find
\be
\d_1\langle \O_{\bar p_1}^{ren} \O_{\bar p_2}^{ren} \O_{\bar p_3}^{ren}\rangle_{\ln \bar p} 
= \frac{\mu}{2\pi} \left(q_1 \bar p_1 \ln \frac{\bar p_1}{\bar p_2} + q_3 \bar p_3 \ln \frac{\bar p_2}{\bar p_3}\right)\langle  \O_{\bar p_1} \O_{\bar p_2} \O_{\bar p_3}\rangle_{CFT}
\ee
which again is non-vanishing. Note this correction  agrees with \eqref{logtoo} if $q_3=0$.

\subsection{Operator mixing and basis diagonalization  \label{opmix}}

In this subsection, we bring  supporting evidence for the general picture of operator mixing presented in section \ref{genstr}. In particular, we check explicitly  that there are no logarithmic terms in sample off-diagonal correlators, as those would imply a breaking of the operator degeneracy. 
 We pay special attention to the mixing of the current and its Virasoro$_R$ descendants. 

\subsubsection*{Mixing between $\bar T \O $ and $\O$}

As explained in section \eqref{genstr}, in order to study operator mixing we need to first decompose the original Virasoro$_R$ representations into global $SL(2,\mathbb{R})_R$ ones. At level zero, we  have the primary operator $\O$, whereas at level two, there is a new $SL(2,\mathbb{R})_R$  primary

\be
(\bar T \O) = :\! \bar T \O \!: - \frac{3}{2(2\bar h+1)} \bar \p^2 \O
\ee
The two-point function of this operator is

\be
\langle (\bar T \O)_1 (\bar T \O)^\dag_2 \rangle = \frac{\mathcal{N}_{\bar T \O}}{z_{12}^{2h}\bar z_{12}^{2\bar h+4}}\;, \;\;\;\;\;\;\; \mathcal{N}_{\bar T \O} = \frac{c}{2}+ \frac{\bar h (8\bar h-5)}{2\bar h+1} \label{normto}
\ee
and $(\bar T \O)$ is orthogonal to $\O$. At linear order in $\mu$, the two operators mix 

%
\be
\d_1 \langle (\bar T \O)_1  \O_2^\dag\rangle  =\mu_E  \langle (\bar T \O)_1 \O_2^\dag \int d^2 z_3 \, : \! J \bar T_3\!:\rangle = \frac{q \mu_E\, \mathcal{N}_{\bar T \O}}{ z_{12}^{2h-1}\bar z_{12}^{2\bar h} } \int \frac{d^2 z_3}{ z_{13} z_{23}\bar z_{13}^4} = \frac{\frac{2\pi q}{3} \mu_E \mathcal{N}_{\bar T \O}}{z_{12}^{2h} \bar z_{12}^{2\bar h+3} }
\ee
The integral does not produce a log, which implies that the matrix $N$ in \eqref{mnmat} does not acquire an off-diagonal term, in agreement with our prediction that operators will stay degenerate to $\O(\mu)$. To show that $(\bar T \O)$ and $\O$ stay degenerate to linear order, we compute the anomalous dimension  of $(\bar T \O)$ 

\be
\d_1 \langle (\bar T \O)_1 (\bar T \O_2)^\dag\rangle  =\mu_E \int d^2 z_3 \langle (\bar T \O)_1 (\bar T \O_2)^\dag :\!J \bar T_3\!:\rangle =\mu_E \int d^2 z_3  \frac{q(\bar h +2)\, \mathcal{N}_{\bar T \O}}{z_{12}^{2h-1} z_{13} z_{23}\, \bar z_{12}^{2\bar h +2} \bar z_{13}^2 \bar z_{23}^2}
\ee
Comparing with \eqref{intanomd}, we note that the integrand is 
 identical to that of an operator of dimension $\bar h+2$ with the  unconventional normalization \eqref{normto}. Consequently, upon Fourier transform 
 the anomalous dimension will be the same as that of $\O$.   In order to make the $\{ \O, (\bar T \O), \ldots \}$ basis diagonal to order $\mu$, we simply need to redefine

\be
(\bar T \O) \r (\bar T \O) + \frac{\pi  q \mu_E \, \mathcal{N}_{\bar T \O}}{6 \bar h (\bar h+1) (2 \bar h+1)} \bar \p^3 \O
\ee
Since $\O$ and $(\bar T \O)$ are orthogonal in the undeformed CFT, the anomalous dimension of $(\bar T \O)$ is unaffected by this shift. 

This analysis can be in principle continued to arbitrarily high order. It would be interesting to have an all-orders proof that the structure we obtain is the one described in section \eqref{cptsetup}.  


\subsubsection*{Mixing between $\bar T$ and $ \bar T^2 $}

We  now  repeat the above analysis  for 
%
the operators starting off as $\bar T$ and $(\bar T)^2$, both of which are  Virasoro$_R$ descendants of the identity operator. In particular, we show that these operators follow exactly the same mixing pattern as generic operators in the deformed CFT.

 While $\bar T$ is an  $SL(2,\mathbb{R})$ primary in the undeformed CFT, the level four primary $(\bar T)^2$ is given by

\be
 (\bar T^2) \equiv\; :\! \bar T\bar T \!:-  \frac{3}{10} \bar \p^2 \bar T  
\ee
The normalization of the $(\bar T^2)$ two-point function is 

\be
\langle (\bar T^2)_1 (\bar T^2)_2 \rangle = \frac{\mathcal{N}_{\bar T^2}}{\bar z_{12}^8}\;, \;\;\;\;\;\;\;\; \mathcal{N}_{\bar T^2}= \frac{c(5c+22)}{10 }
\ee
Let us now compute the mixing between $\bar T$ and $(\bar T^2)$, which receives the first non-trivial contribution at $\O(\mu^2)$

\be
\d_2 \langle (\bar T^2)_1 \bar T_2 \rangle = \frac{\mu_E^2}{2}\int d^2 z_3 d^2 z_4 \left\langle \left(:\!\bar T \bar T_1\! :- \frac{3}{10} \bar \p_1^2 \bar T_1 \right)\bar T_2 : \!J \bar T_3\!: : \!J \bar T_4\!: \right\rangle
\ee
The above correlator of stress tensors takes a surprisingly simple form, 
since the $\bar z_i$ dependence of the integrand is the same in the coefficient proportional to $c$ as in that proportional to $c^2$. More precisely, we have
\be
\d_2 \langle (\bar T^2)_1 \bar T_2 \rangle = \frac{c(c+\frac{22}{5})k \mu_E^2}{4}\int \frac{d^2 z_3 d^2 z_4}{z_{34}^2} \left(\frac{1}{\bar z_{12}^4 \bar z_{13}^2 \bar z_{14}^2 \bar z_{34}^2} + \frac{1}{\bar z_{13}^4 \bar z_{12}^2 \bar z_{14}^2 \bar z_{24}^2} + \frac{1}{\bar z_{14}^4 \bar z_{12}^2 \bar z_{13}^2 \bar z_{23}^2} \right) 
\ee
These integrals do not give rise to logarithms, as expected. The polynomial term can be absorbed into a redefinition of $(\bar T^2)$ by $\mu^2 \bar \p^4 \bar T$, 
which is precisely of the form  \eqref{lowerdiag} we argued for. This confirms that at least to this order, $\bar T$ follows the same mixing pattern as generic operators in the CFT. Note however that, as already mentioned in section \ref{cptsetup}, the  operator $\bar T_{\bar p}$ studied here is not the same as the  Noether current associated to right-moving translations.




\subsubsection*{Mixing between $J$ and $J\bar T$}

The argument of section \ref{cptsetup}, which we checked above in examples, would lead us to think that we never need to consider mixing between $\O$ and higher powers of $\bar T^n \O$ when diagonalizing the operator basis.  
%
While this is generally true, there exist exceptions  if the operator in question is purely holomorphic, as for example in the case of the chiral current $J$. 

The two-point function of the current first receives corrections at $\O(\mu^2)$, but the correction is a pure contact term\footnote{It is interesting to compare this calculation with its counterpart in the $J\bar J$ deformation case, where the correction to the level is finite. }
\be
\d_2 \langle J_1 J_2 \rangle = \frac{\mu_E^2}{2}  \langle J_1 J_2 \int d^2 z_3 :\!J \bar T_3 \!: \!\int d^2 z_4 :\! J \bar T_4\!: \rangle 
= - \frac{\pi^3 \mu_E^2 k^2 c}{6} \p_1 \bar \p_1 \, \d(z_{12})\d(\bar z_{12})
\ee
which we ignore (it can also be removed by a shift of $J$ by $\mu^2 \bar \p^2 J$). Consequently, $J$ does not acquire an anomalous dimension to this order, which is consistent with the fact that it has $\bar p_J =0$. 

The current $J$ does mix non-trivially with $J\bar T$ at second order in $\mu$, as can be seen by computing

\be
\d_2 \langle J_1 \!:\!J \bar T\!_2:  \rangle =\frac{\mu_E^2}{2} \langle J_1 :\! J \bar T_2 \!:\int d^2 z_3 :\!J \bar T\!_3: \int d^2 z_4 :\!J \bar T_4\!: \rangle
\ee
with the result

\be
\d_2 \langle J_1 :\! J \bar T_2\!:  \rangle =\frac{\mu_E^2 k^2 c}{8} \int \frac{d^2 z_3 d^2 z_4}{\bar z_{23}^2 \bar z_{34}^2 \bar z_{24}^2} \left(\frac{1}{z_{12}^2 z_{34}^2} + \frac{1}{z_{13}^2 z_{24}^2} + \frac{1}{z_{14}^2 z_{23}^2}\right) = \frac{\mu_E^2 \pi^2 k^2 c}{2 z_{12}^2 \bar z_{12}^4} \label{corrjjtbar}
\ee 
Note that we cannot reset this off-diagonal term to zero by shifting $J\bar T$ by $\mu^2 \bar \p^4 J$, because  the correlation function of the shift with $J$ is a pure contact term in the original CFT. Thus, the only way to render the basis diagonal is by instead redefining $J$ as 

\be
J \r J'=  J - 2 \mu_E^2 \pi^2 k :J\bar T: + \ldots \label{jredefmusq}
\ee
The same shift in the current was obtained in \cite{kutasov}, by requiring that the current stay chiral in the deformed theory. Indeed, acting with $\p_{z_1}$ on \eqref{corrjjtbar} yields a non-zero answer, which means that the current was no longer holomorphic before we shifted it. 


The  $J\bar T$ operator does, on the other hand, behave as a generic operator from the tower \eqref{otower}, as one expects from the fact that it is not holomorphic.  Its two-point function is first corrected at $\O(\mu^2)$

\be
\d_2 \langle :\! J\bar T_1 \!: :\! J \bar T_2 \!: \rangle = \frac{\mu_E^2 k^2}{8} \int d^2 z_3 d^2 z_4 \left( \frac{1}{z_{12}^2 z_{34}^2} + \frac{1}{z_{13}^2 z_{24}^2} + \frac{1}{z_{14}^2 z_{23}^2} \right) \langle \bar T_1 \bar T_2 \bar T_3 \bar T_4 \rangle
\ee
The integral over the first term is identical to the one performed in \eqref{del2tbarsq} and  produces the correct coefficient of the logarhitmic divergence to correspond to  an anomalous dimension $\mu^2 k \bar p^2/16\pi^2 $. The integrals over the other two terms are equal to each other and we need to check that they do not lead to any new logarhitmic divergences. This is easily  checked for the terms proportional to $c^2$ in \eqref{stresst4pf}. We are left to evaluate 

\be
2 \cdot \frac{\mu_E^2 k^2 c}{8} \int \frac{d^2 z_3 d^2 z_4}{z_{13}^2 z_{24}^2} \left(\frac{1}{\bar z_{12}^2 \bar z_{13}^2 \bar z_{24}^2 \bar z_{34}^2} +  \frac{1}{\bar z_{12}^2 \bar z_{14}^2 \bar z_{23}^2 \bar z_{34}^2} + \frac{1}{\bar z_{13}^2 \bar z_{14}^2\bar  z_{24}^2\bar z_{23}^2}  \right)
\ee
Even though the first two integrals are logarithmically divergent, their divergences cancel exactly against each other. 

The conclusion of this analysis is that in order to keep a diagonal operator basis, $J$ needs to be corrected at $\O(\mu^2)$ as in \eqref{jredefmusq}. The form of the correction  is a consequence of the holomorphy of the current. Higher order corrections of the form $\mu^{2m} J (\bar T^m)$ would also be necessary if the corresponding $\d_{2m} \langle J :\! J \bar T^m\!\! : \rangle$ matrix elements are non-zero.  One the other hand, since the operator $J\bar T$ is not holomorphic, its matrix elements with the other $J\bar T^n$ can be rediagonalized via a basis change of the form \eqref{lowerdiag}, i.e. it is only  (multiplicatively) renormalized by derivatives of itself.
%
%
%
%
%
%
%
%
%
 It would be interesting to better understand the relation between this operator  and the one defined via the OPE of the deformed currents. 

\subsubsection*{Mixing between $\bar T$ and $J\bar T$}

This is again an interesting case of mixing. 
Already at linear order in $\mu$, we find a non-zero answer

\be
\d_1\langle \bar T_1 :\!J \bar T_2\!: \rangle = \mu_E \langle \bar T_1 :\!J \bar T_2\!: \int d^2 z_3 : \!J\bar T_3 \!:\rangle = \frac{\mu_E k c}{2 \bar z_{12}^2} \int \frac{d^2 z_3}{z_{23}^2 \bar z_{23}^2 \bar z_{13}^2}
= - \frac{2 \pi \mu_E k c}{z_{12} \bar z_{12}^5} \label{jjtbarcorr}
\ee
Since the $\p_{z_1}$ derivative of the above does not vanish, it reflects the fact that the stress tensor is no longer holomorphic at first order in $\mu_E$. This agrees with the results of \cite{kutasov}, where $\p \bar T =  - 2\pi \mu_E \bar \p J \bar T$. 

Naively, \eqref{jjtbarcorr} suggests that a \emph{non-local} redefinition of the stress tensor, such as \eqref{nonloccorr}, is needed in order to keep the basis diagonal.
However, such a redefinition is not allowed within our framework, where locality in the $z$ direction should always be manifest. Additionally, it is easy to check that such a shift would affect the anomalous dimension of $\bar T$, which would no longer match \eqref{hqtbar}.  

 An interpretation of this non-zero matrix element that does fit within our framework is the following. As mentioned in section \ref{genstr}, the operators in the $J\bar T$-deformed CFT belong to  left-moving Virasoro-Ka\v{c}-Moody representations. The Ka\v{c}-Moody descendants may be written in terms of Virasoro$_L$ primaries, e.g. at level one we have

\be
(J\O) = :J\O: - \frac{q}{2h} \p \O
\ee
which has vanishing inner product with $\O$. Note that now the bracket notation stands for a Virasoro, rather than a global, primary. 
For the case of the stress tensor, the corresponding combination in the deformed theory  should be
\be
(J \bar T) = :J\bar T: - \frac{q_{\bar T}}{2h_{\bar T}} \p \bar T =  :J\bar T: - \frac{2\pi}{\mu \bar p}  \p \bar T
\ee
Thus,  the expansion of this operator around $\mu =0$ looks singular, if we choose  the coefficient of the $:\!J\O\!:$ term to be one. By our general recipe, this primary operator should be orthogonal to the primary constructed from $\bar T$. 
It is not hard to check that this is inded the case: if we compute $\langle \p \bar T \bar T \rangle $ in conformal perturbation theory using \eqref{tbar2pfmusq}, the first non-vanishing contribution is at second order in $\mu$, which perfectly cancels the contribution \eqref{jjtbarcorr} from  the first term.  We conclude that  the non-zero two-point function of $J\bar T$ with $\bar T$ is simply due to the fact that $J\bar T$ is not a Virasoro primary, but it needs to be shifted by  $\p \bar T$ in order to become one, with a coefficient that is divergent as $\mu \r 0$. 
If we do not mind working with Virasoro descendants, then these singular-looking combinations need not be considered. 
 

\section{Discussion \label{disc}}

In this article, we have made the first  steps towards the specification of  correlation functions in $J\bar T$-deformed CFTs in terms of the conformal data of the original CFT. In particular, we have proposed an exact formula 
for the spectrum of conformal dimensions and charges of $J\bar T$-deformed CFTs and checked it to leading order in conformal perturbation theory. We have also computed the OPE coefficients to linear order in the perturbation and 
addressed the issue of operator mixing.

To make further progress, there are several technical and conceptual issues that  need to be resolved. First, one needs a better understanding of the deforming operator, and in particular of the two currents that compose it. While the chiral  current $J$ is well-defined and relatively straightforward to construct order by order in conformal perturbation theory, it is not completely clear what the definition of $\bar T$ should be in the deformed CFT. As we pointed out, the CFT antiholomorphic stress tensor gives rise to at least two different notions of $\bar T$ operator in the deformed theory: the primary operator $\bar T_{\bar p}$ discussed in this article, with anomalous dimension and charge given by \eqref{hqtbar},  and the $\bar z$ component of the Noether current associated to right-moving translations, which 
depends on the non-local combination $\bar z - \mu \int J(z)$ and does not yet have a  definition outside the Lagrangian framework. It would be interesting to understand how these two notions relate to each other (even though the difference between them is irrelevant at the level of the integrated operator, which projects on their coinciding $\bar p =0$ component).
%
 It would also be interesting to understand the difference between the two  notions of  $J\bar T$ operator that we discussed. 
 %
 %
%
Another issue that one should  address is whether contact terms, which we have completely ignored in our analysis, may play a role in computing the deformed correlators. Finally, it would be very interesting if one could obtain flow equations for the  correlation functions  as $\mu$ is varied, along the lines  \cite{Aharony:2018vux} followed for the case of $T\bar T$, which may allow us to access the correlation functions at finite $\mu$.

Another important technical  issue concerns the choice of ultraviolet regulator for the divergences of the integrated correlators. Usually, the UV regulator is chosen to respect as many symmetries of the problem as possible. Our choice of   a
standard Lorentz invariant  regulator, however, does not present any particular advantage, since  the symmetry  it is designed to respect is explicitly broken by the deformation. While it is encouraging that the one-dimensional conformal structure of the deformed correlators agrees with the general expectations, the rather unappealing form of the OPE coefficients we have found is a direct consequence of having used a Lorentz-invariant ultraviolet regulator.  We cannot help wondering whether  a different  regulator may be more natural in $J\bar T$-deformed CFTs, especially since these theories do not posses a usual UV fixed point as far as the right-movers are concerned, and may be sensitive to the particular choice we make. The question of suitable  ultraviolet regulators  can in principle also be studied in the context of dipole theories \cite{Bergman:2000cw}, whose UV structure is expected to be similar to that of $J\bar T$-deformed CFTs and for which Lagrangian and integrability-based methods are also available \cite{Guica:2017mtd}.
Apart from these ultraviolet issues, note that  one also needs to worry about possible infrared problems due to
%
 the presence of a continuous spectrum of operators, which extends all the way to zero dimension.

%
%
%

Once these issues are resolved, there are many interesting features of $J\bar T$-deformed CFTs that one can explore, and possible new structures to uncover. For example,  if it is found that the OPE coefficients do not receive corrections in perturbation theory, then    $J\bar T$-deformed CFTs can be interpreted in terms of   momentum-dependent spectral flow and operator-dependent coordinate transformations. These  would 
not only provide a more geometric picture for the deformation directly in right-moving position space, but they  would also allow one to specify all correlation functions in $J\bar T$-deformed CFTs by applying a simple operation to the correlation functions of the original CFT.

A very interesting question is whether $J\bar T$-deformed CFTs possess  hidden symmetries associated to the right-moving Virasoro symmetry of the original CFT. There are two different indications that this structure may continue to exist: first, the asymptotic symmetry analysis of \cite{cssint} found that the right-moving translational symmetry $U(1)_R$ was enhanced to  a full right-moving Virasoro symmetry that was field-dependent; second, the analysis of the current paper shows that operators that were originally part of the same Virasoro$_R$ highest weight representation acquire the same anomalous dimension in the deformed theory at fixed $\bar p$, and thus stay degenerate. It would be very interesting to investigate whether these degenerate operators could still be related by  a larger symmetry. For this, one needs to understand how the OPE coefficients of the various $SL(2,\mathbb{R})_R$ primaries in \eqref{otower} are modified relative to each other. 
A complementary approach would be to understand whether a current $\bar T_{Noether}$ associated to the right-moving symmetries can still be defined in the deformed CFT. 

A natural question is whether the results of this paper can be reproduced from a holographic calculation. For this, one needs to extend the holographic dictionary proposed in \cite{cssint}, which was restricted to the study of the currents, to include general propagating degrees of freedom. In particular, it would be interesting to reproduce the shift \eqref{defspec} in the operator dimensions and charges from holography, check our prediction for the OPE coefficients and check how the primary operator $\bar T_{\bar p}$ we discussed fits within the holographic framework.

Finally, let us note that $J\bar T$-deformed CFTs provide the first concrete example of a dipole CFT$_2$,  a type of quantum field theory loosely defined as a deformation of a two-dimensional CFT by a set of irrelevant operators of dimensions $(1,n)$, whose coefficients are finely tuned so that the resulting theory is UV-complete, though non-local. That many such dipole CFTs should exist is indicated by the ubiquity of warped AdS$_3$ backgrounds - the spacetimes holographically dual to such field theories - in string theory \cite{Bobev:2011qx,ElShowk:2011cm,Kraus:2011pf}, but so far no general field-theoretical procedure for defining them has been proposed. On the other hand  their study, as argued in \cite{ElShowk:2011cm}, is directly relevant to the holographic description of general extremal black holes \cite{Guica:2008mu}. 
%

A central ingredient in the microscopic description of extremal black holes is the presence of a right-moving Virasoro symmetry, and our hope is that the study of $J\bar T$-deformed CFTs can shed light on this interesting issue.  
 %
One  simple calculation suggested by our analysis is to study operator degeneracy in dipole CFTs and determine whether the original degeneracy associated to the right-moving Virasoro symmetry  is lifted by the deformation. If it is not, then this may have interesting  implications for the survival of a right-moving Virasoro symmetry. More generally, one can try to set up a bootstrap programme for dipole CFTs, which should hold order by order in the deformation parameter, and in which  $J\bar T$-deformed CFTs could be used  as a simple concrete example. One can hope that such a programme would allow one to classify which CFTs admit a dipole deformation and  find the general properties of the resulting spectra.

\subsection*{Acknowledgements}

The author would like to thank Camille Aron, Costas Bachas, Brando Bellazzini, Pierre Heidmann, David Kutasov,  Ruben Monten, Miguel Paulos, Slava Rychkov, Andrew Strominger, Marika Taylor, Jan Troost and  Xi Yin  for interesting discussions, and  Adam Bzowski for collaboration in the early stages of this project. This research was supported in part by the National Science Foundation under Grant No. NSF PHY-1748958, as well as by the ERC Starting Grant 679278 Emergent-BH and the Swedish Research Council grant number 2015-05333.

\appendix




\section{List of basic correlators \label{listcorr}}

In this appendix, for easy reference,  we collect the list of basic correlators that are used in the main text. They usually involve a primary operator $\O$ of dimension $(h,\bar h)$ and charge $q$ and/or the currents $J, T$ and $\bar T$. The notation $\O_i$ stands for $\O_i(z_i,\bar z_i)$. 

The correlation functions that involve the currents are easily determined by repeated use of the Ward identities
\bea
\langle J(z) \, \O_1 \ldots \O_n J_{n+1} \ldots J_{n+m} \rangle & = &  \sum_{i=1}^n \frac{q_i}{z-z_i} \langle  \O_1 \ldots \O_n J_{n+1} \ldots J_{n+m} \rangle + \nonumber \\
& & \hspace{-3cm} + \sum_{j=1}^m \frac{k/2}{(z-z_{n+j})^2 } \langle  \O_1 \ldots \O_n J_{n+1} \ldots \check{J}_{n+j} \ldots J_{n+m} \rangle
\eea
where the notation $\check{J}$ means that the corresponding term has been ommitted. The stress tensor Ward identity reads
\bea
\langle T(z)\, \O_1 \ldots \O_n T_{n+1} \ldots T_{n+m} \rangle & = &  \sum_{i=1}^{m+n} \left( \frac{h_i}{(z-z_i)^2}+ \frac{1}{z-z_i} \p_{z_i}\right) \langle  \O_1 \ldots \O_n T_{n+1} \ldots T_{n+m} \rangle + \nonumber \\
& & \hspace{-3cm} + \sum_{j=1}^m \frac{c/2}{(z-z_{n+j})^4 } \langle  \O_1 \ldots \O_n T_{n+1} \ldots \check{T}_{n+j} \ldots T_{n+m} \rangle
\eea
where in the first sum $h_i=2$ for $i>n$.  
The Ward identity for the antiholomorphic stress tensor is identical, upon the replacements $T \r \bar T$,  $h \r \bar h$ and $z \r \bar z$.

\subsubsection*{Correlation functions involving a generic operator}

 The two-point function of a generic  primary operator $\O$ of dimension $(h,\bar h)$ and charge $q$ is

\be
\langle \O_1 \O^\dag_2 \rangle = \frac{1}{z_{12}^{2h} } \cdot \frac{1}{ \bar{z}_{12}^{2\hbar}} 
\ee
Using the above Ward identities, we obtain

\be
\langle \O_1  \O^\dag_2 J_3 \bar T_3 \rangle = \frac{q }{z_{12}^{2h-1} z_{13} z_{23} } \cdot \frac{\bar h}{\bar{z}_{12}^{2\hbar-2} \bar{z}_{13}^2 \bar z_{23}^2}
\ee

\be
\langle \O_1 \O^\dag_2 J_3 J_4 \bar T_4 \rangle = \left(\frac{k}{z_{34}^2} + \frac{(q\, z_{12})^2}{ z_{13} z_{23} z_{14} z_{24}} \right)\frac{1}{z_{12}^{2h} } \cdot \frac{\bar h}{\bar{z}_{12}^{2\hbar-2} \bar{z}_{14}^2 \bar z_{24}^2}
\ee

\be
\langle \O_1 \O^\dag_2  T_3 J_4 \bar T_4 \rangle = \left(\frac{h z_{12}^2}{z_{13}^2 z_{23}^2} + \frac{z_{14} z_{24}}{z_{13} z_{23} z_{34}^2} \right) \frac{q }{z_{12}^{2h-1} z_{14} z_{24} } \cdot \frac{\bar h}{\bar{z}_{12}^{2\hbar-2} \bar{z}_{14}^2 \bar z_{24}^2}
\ee

\be
\langle \O_1 \O^\dag_2 \bar T_3 J_4 \bar T_4 \rangle = \frac{q }{z_{12}^{2h-1} z_{14} z_{24} } \cdot \frac{1}{\bar{z}_{12}^{2\hbar}}\left(\frac{\bar h^2 \bar z_{12}^4}{\bar z_{13}^2 \bar z_{23}^2 \bar{z}_{14}^2 \bar z_{24}^2} + \frac{2 \bar h \bar z_{12}^2}{\bar z_{14} \bar z_{24} \bar z_{13} \bar z_{23} \bar z_{34}^2} +\frac{c}{2 \bar z_{34}^4} \right) 
\ee

\be
\langle \O_1 \O^\dag_2  J_3 \bar T_3 J_4 \bar T_4 \rangle = \left(\frac{k}{z_{34}^2} + \frac{(q\, z_{12})^2}{ z_{13} z_{23} z_{14} z_{24}} \right)\frac{1}{z_{12}^{2h} } \cdot  \left(\frac{\bar h^2 \bar z_{12}^4}{\bar z_{13}^2 \bar z_{23}^2 \bar{z}_{14}^2 \bar z_{24}^2} + \frac{2 \bar h \bar z_{12}^2}{\bar z_{14} \bar z_{24} \bar z_{13} \bar z_{23} \bar z_{34}^2} +\frac{c}{2 \bar z_{34}^4} \right) \cdot \frac{1}{\bar{z}_{12}^{2\hbar}} \label{omu2}
\ee

\be
\langle \O_1 \O_2 \O_3 J_4 \bar T_4 \rangle  =\left( \frac{q_1}{z_{41}}+ \frac{q_2}{z_{42}}+\frac{q_3}{z_{43}}\right)  \left(\frac{\bar h_1 \bar z_{12}\bar z_{13}}{\bar z_{14}^2 \bar z_{24} \bar z_{34}}-\frac{\bar h_2 \bar z_{12}\bar z_{23}}{\bar z_{14} \bar z_{24}^2 \bar z_{34}}+\frac{\bar h_3 \bar z_{13}\bar z_{23}}{\bar z_{14} \bar z_{24} \bar z_{34}^2} \right) \langle \O_1 \O_2 \O_3 \rangle
\ee

\subsubsection*{Correlation functions of the currents}

Correlation functions of $J$'s are simply given by Wick contractions of the basic correlator 

\be
\langle J_1 J_2 \rangle = \frac{\frac{k}{2}}{z_{12}^2} 
\ee
The first few correlation functions of the stress tensor are given by

\be
\langle \bar T_1 \bar T_2 \rangle = \frac{c/2}{\bar z_{12}^4} \;, \;\;\;\;\;\;\;\langle \bar T_1 \bar T_2 \bar T_3\rangle = \frac{c}{\bar z_{12}^2\bar z_{13}^2 \bar z_{23}^2} 
\ee
The four-point function of four stress tensors is

\be
\langle\bar T_1 \bar T_2 \bar T_3 \bar T_4 \rangle = \frac{c^2}{4} \left(\frac{1}{\bar z_{12}^4\bar z_{34}^4} + \frac{1}{\bar z_{13}^4 \bar z_{24}^4} + \frac{1}{\bar z_{23}^4\bar z_{14}^4} \right) + c \left(\frac{1}{\bar z_{12}^2 \bar z_{13}^2 \bar z_{24}^2 \bar z_{34}^2} +  \frac{1}{\bar z_{12}^2 \bar z_{14}^2 \bar z_{23}^2 \bar z_{34}^2} + \frac{1}{\bar z_{13}^2 \bar z_{14}^2\bar  z_{24}^2\bar z_{23}^2}  \right) \label{stresst4pf}
\ee
Correlation functions of double-trace operators $:\!AB\!:$ are obtained by keeping the $\O(z^0)$ term in the expansion of $\langle A(z)B(0)\ldots \rangle$.

%
%
%

%
%
%
%
%
%

\section{Useful  integrals \label{useint}}

There are two basic integrals that appear recurrently in our calculations.  

\subsubsection*{The $\mathcal{I}$ integral}

\be
\mathcal{I}(z_1,z_2) \equiv \int   \frac{d^2 z_3}{|z_{13}|^2 |z_{23}|^2}
\ee
The integral can be performed by introducing a Schwinger parameter
\be
\mathcal{I}(z_1,z_2)= \int_0^1 du \int \frac{d^2 z_3}{(u |z_{13}|^2 + (1-u) |z_{23}|^2)^2} = \int_0^1 du \int \frac{d^2 z'_3}{(|z_3'|^2 + u(1-u) |z_{12}|^2)^2} 
\ee
Since the result is divergent,  we  evaluate it using dimensional regularization, i.e.  we replace $d^2 z_3 \r d^d z_3$. Then
\bea
\mathcal{I}(z_1,z_2) &=& 2 V_{S^{d-1}} \int_0^1 du \int_0^\infty \frac{\rho^{d-1} d\rho}{(\rho^2 + u (1-u)|z_{12}|^2)^2}  \nonumber \\
&=&V_{S^{d-1}} \Gamma\left(\frac{d}{2}\right) \Gamma\left(2-\frac{d}{2}\right) |z_{12}|^{d-4} \int_0^1 du \, [u(1-u)]^{\frac{d}{2}-2} \nonumber \\
&=& 2\pi^{\frac{d}{2}}\Gamma\left(2-\frac{d}{2}\right) \mathrm{B}\left(\frac{d}{2}-1,\frac{d}{2}-1\right)\,  |z_{12}|^{d-4}
\eea
where  the volume of $S^{d-1}$ is $V_{S^{d-1}} =  2 \pi^{d/2}/\Gamma(d/2)$. Taking $d= 2 + \e$ and expanding 
we find

\be
\mathcal{I}(z_1,z_2) = 2\frac{2\pi}{|z_{12}|^2} \left( \frac{2}{\e} + \ln |z_{12}|^2 + \g + \ln \pi  + \frac{\e}{4} (\ln |z_{12}|^2 + \g +\ln \pi)^2 - \frac{\e \pi^2}{24} + \O(\e^2)\right) \label{iint}
\ee
In the main text, we will only need the expansion of this integral up to $\O(\e)$. 

\subsubsection*{The $I_{ij}$ integral}

Another useful integral is
\be
I_{ij} \equiv \int \frac{d^2 z_4}{z_{4i} \bar z_{4j}} 
\ee
For definiteness, we compute $I_{12}$

\be
I_{12}= \int \frac{d^2 z_4\, \bar z_{41} z_{42}}{|z_{41}|^2 | z_{42}|^2}= \int_0^1 du \int \frac{d^2 z_4 \,  \bar z_{41} z_{42}}{[u |z_{41}|^2 + (1-u)|z_{42}|^2]^2}= \int_0^1 du \int \frac{d^2 z'_4 \,  (\bar z_{4}' -(1-u) \bar z_{12}) (z_{4}'+u z_{12})}{[z_4'^2 + u(1-u) |z_{12}|^2]^2} \nonumber
\ee
Changing again the dimension to $d$ we find
\bea
I_{12}&=& 2 V_{S^{d-1}}\int_0^1 du \int_0^\infty  d\rho \frac{ \rho^{d-1} (\rho^2 -u(1-u) |z_{12}|^2)}{(\rho^2 + u(1-u)|z_{12}|^2)^2} = 2 \frac{2\pi^{d/2}}{\Gamma(\frac{d}{2})} \int_0^1 du \frac{ (d-1)[u(1-u)|z_{12}|^2]^{\frac{d}{2}-1}}{2} \Gamma(\frac{d}{2}) \Gamma(1-\frac{d}{2}) \nonumber \\
&& =2\pi^{d/2} |z_{12}|^{d-2} \frac{\Gamma(d/2)^2 \Gamma(1-d/2)}{\Gamma(d-1)} = -2\pi \left(\frac{2}{\e} + \ln |z_{12}|^2 + \g_E + \ln \pi  + \O(\e) \right)
\eea

\subsubsection*{Derivatives of $I_{ij}$ }

Differentiating this result, we can obtain a list of simple integrals 

\be
\int \frac{d^2 z_4}{z_{43}^2 \bar z_{41}} = \frac{2\pi}{z_{13}}\;, \;\;\;\;\;\;\; 
\int \frac{d^2 z_4}{z_{43}^2 \bar z_{41}^2} = 4 \pi^2 \d^2(z_{13})
\ee
where we use $\bar \p \frac{ 1}{z} =  \p \frac{ 1}{\bar z} = 2 \pi \d(z) \d(\bar z)$. All integrals in the main text are computed by taking differences and derivatives of this basic integral. For example, the integral appearing in \eqref{jooint} is calculated as

\be
 \int \frac{d^2 z_4}{z_{34}^2 \bar{z}_{14}^2 \bar z_{24}^2}= \p_{\bar z_1} \p_{\bar z_2} \frac{1}{\bar z_{12}}\int  \frac{d^2 z_4 }{z_{34}^2}  \left( \frac{1}{\bar z_{24}}-\frac{1}{ \bar{z}_{14}} \right) =- \frac{4\pi}{\bar{z}_{12}^3} \left(\frac{1}{z_{13}}- \frac{1}{z_{23}} \right) \label{joointev}
\ee
It is also easy to check that $\mathcal{I}(z_1,z_2) = \frac{-2}{|z_{12}|^2} I_{12}$, without needing to evaluate the integrals.

\section{Fourier transforms \label{fourier}}

\subsubsection*{Basic Fourier transforms}

We start with the Fourier transform of expressions of the type $\bar z^{-a}$ and $\bar z^{-a} \ln |\bar z|$, where $a$ is an arbitrary real number (the case of $a$ integer sometimes needs separate treatment) . We mostly follow the treatment in \cite{fourier}. To compute the Fourier transform of $\bar z^{-a}$, we consider instead $\frac{1}{\bar z^a} \Theta (\bar z)$. We further regulate it as

\be
\mathcal{F}_{t} \left( \frac{\Theta(\bar z)}{\bar z^{a}} \right)=\lim_{t \r 0} \int_0^\infty \frac{d\bar z}{ \bar z^{a}}\, e^{- \bar z (t+i \bar p)} = \lim_{t \r 0}  (t+i \bar p)^{a-1} \Gamma(1-a) = \frac{\pi e^{\frac{i \pi }{2}  (a-1) sgn \bar p} |\bar p|^{a-1}}{\sin \pi a \, \Gamma(a)}
\ee
To find the Fourier transform of $\bar z^{-a}$ alone we must evaluate

\be
\int_{-\infty}^0  \frac{d\bar z}{ \bar z^{a}}\, e^{- i \bar z  \bar p}= (-1)^{-a} \int_0^\infty \frac{d\bar z}{ \bar z^{a}}\, e^{i \bar z  \bar p}
\ee
Notice that $\bar z^{-a}$ has a branch cut along the real negative axis, and we must specify where we place it. We choose to place it below the real axis by the replacement $\bar z \r \bar z + i \varepsilon$. In that case

\be
\int_{-\infty}^\infty \frac{d\bar z}{ \bar z^{a}}\, e^{- i \bar z  \bar p} = (e^{\frac{i \pi }{2}  (a-1) sgn \bar p}+ e^{-i \pi a} e^{-\frac{i \pi }{2}  (a-1) sgn \bar p}) \frac{\pi  |\bar p|^{a-1}}{\sin \pi a \, \Gamma(a)} = \frac{2\pi (-i)^{a} |\bar p|^{a-1}}{\Gamma(a)} \Theta (\bar p)
\ee
The fact that the Fourier transform of the Wightman function vanishes for negative frequencies is  well known, which follows from the fact that for $\bar p <0$ we should close the integration contour in the upper half plane. Since the branch cut is below the real axis, the integral can be deformed to zero.  Thus, we have
\be
\mathcal{F}_{t} \left( \frac{1}{\bar z^{a}} \right)= 2\pi  \frac{(-i)^{a} |\bar p|^{a-1}}{\Gamma(a)}\, \Theta (\bar p)
\ee
The Fourier transform of $\bar z^{-a} \ln \bar z$ can be obtained by differentiation with respect to $a$, obtaining

\be  \mathcal{F}_{t} \left( 
\frac{\ln \bar z}{\bar z^{a}} \right)= 2\pi \frac{(-i)^{a} |\bar p|^{a-1}}{\Gamma(a)}  \, \left( \psi (a)-\ln |\bar p| + \frac{i \pi}{2}\right) \Theta(\bar p)
\ee

\subsubsection*{Fourier transform of the three-point function}
The Fourier transform of the three-point function reads
\be
K(a,b,c)\, \d(\bar p_1 + \bar p_2 + \bar p_3) = \int \prod_{k=1}^3 dz_k \frac{e^{-i \bar p_k \bar z_k}}{\bar z_{12}^a \bar z_{23}^b \bar z_{13}^c} = \int dx dy \frac{e^{-i\bar p_1 x + i \bar p_3 y}}{x^a y^b (x+y)^c} \d(\bar p_1 + \bar p_2 + \bar p_3)
\ee
The integral can be performed by introducing a Schwinger parameter $u$ \cite{volovich}
\bea
K (a,b,c) &=& \frac{\Gamma(a+c)}{\Gamma(a)\Gamma(c)}\int_0^1 du\, u^{a-1} (1-u)^{c-1} \int dx dy \frac{e^{-i\bar p_1 x + i \bar p_3 y}}{(x + (1-u)y)^{a+c} y^b} \\
&=&  \frac{4\pi^2 (-i)^{a+b+c} \Theta(\bar p_1) }{\Gamma(a)\Gamma(b)\Gamma(c)} \times |\bar p_1|^{a+c-1} \int_0^1 du\, u^{a-1} (1-u)^{c-1}  |\bar p_2 + u \bar p_1|^{b-1} \Theta (\bar p_2 + u \bar p_1) \nonumber 
\eea
In the above, we have given $y=z_{23}$ and $x'=z_{12} +(1-u) z_{23}$ a small positive imaginary part, corresponding to an ordering $Im\, t_1 < Im \, t_2 < Im \, t_3$.
The remaining integral is easy to perform if we assume  e.g. that $\bar p_2 >0$ (for $\bar p_2 < - |\bar p_1|$ the integrand vanishes), obtaining\footnote{We have used the integral representation of the hypergeometric function
\be
{}_2 F_1 (a,b,c,z) = \frac{\Gamma(c)}{\Gamma(b) \Gamma (c-b)} \int_0^1 du \, u^{b-1} (1-u)^{c-b-1} (1- u z)^{-a}
\ee}

\be
K (a,b,c)=  \frac{4\pi^2 (-i)^{a+b+c} }{\Gamma(a+c)\Gamma(b)} \, \bar p_1^{a+c-1} \bar p_2^{b-1} \, {}_2 F_1 \left(1-b,a,c+a,-\frac{\bar p_1}{\bar p_2}\right) \label{Kp1p2rep}
\ee
Of course, this result can be written in term of any two of the momenta, using the conservation equation, though its range of validity is for $\bar p_{1,2} >0$.  We can obtain  other representations of the three-point function that are valid in different ranges. For example, if we introduce the Schwinger parameter in a slightly different way, we arrive at
\bea
K (a,b,c) &=& \frac{\Gamma(b+c)}{\Gamma(b)\Gamma(c)}\int_0^1 du\, u^{b-1} (1-u)^{c-1} \int dx dy \frac{e^{-i\bar p_1 x + i \bar p_3 y}}{x^a (y + (1-u)x)^{b+c}}\label{kabcinter} \\
&=&  \frac{4\pi^2 (-i)^{a+b+c} \Theta(-\bar p_3) }{\Gamma(a)\Gamma(b)\Gamma(c)} \times |\bar p_3|^{b+c-1} \int_0^1 du\, u^{b-1} (1-u)^{c-1}  |\bar p_2 + u \bar p_3|^{a-1} \Theta (-\bar p_2 - u \bar p_3) \nonumber 
\eea
If $\bar p_2 <0$, then the theta function can be dropped and the integral evaluates to 

\bea
K (a,b,c) &= &\frac{4\pi^2 (-i)^{a+b+c} |\bar p_3|^{b+c-1}  |\bar p_2|^{a-1}}{\Gamma(a)\Gamma(b+c)} {}_2 F_1 \left(1-a,b,b+c,-\frac{\bar p_3}{\bar p_2}\right)  \nonumber \\
&=& \frac{4\pi^2 (-i)^{a+b+c} |\bar p_3|^{b+c-1}  |\bar p_1|^{a-1}}{\Gamma(a)\Gamma(b+c)} {}_2 F_1 \left(1-a,c,b+c,-\frac{\bar p_3}{\bar p_1}\right) 
\eea
where in the last line we used the  transformation of the hypergeometric,

\be
{}_2 F_1 (a,b,c,z) = (1-z)^{-a} {}_2 F_1 \left(a,c-b,c,\frac{z}{z-1}\right) 
\ee
This expression, which can be obtained by a simple permutation of the arguments and parameters in \eqref{Kp1p2rep},  is valid for $\bar p_{2,3} <0$. It can also be shown that if we insist that $\bar p_2 >0$ in \eqref{kabcinter}, then we are back to the expression  \eqref{Kp1p2rep}. Thus, depending on the signs of the momenta, we obtain different closed-form expressions for $K(a,b,c)$. 

\subsubsection*{Derivatives of the hypergeometric function}

The way we   compute the Fourier transform of power law terms multiplied by  $\ln \bar z_{ij}$ is by taking derivatives of the Fourier transform $K(a,b,c)$  corresponding to the power law  with respect to the parameters $a,b,c$. The expression for $K(a,b,c)$ involves a  hypergeometric function, and taking derivatives with respect to the parameters is usually cumbersome; however, the expression for the hypergeometric \eqref{Kabc} simplifies drastically  when $b$ is  small integer, since the hypergeometric function becomes a polynomial of degree $b-1 \equiv m$
\be
{}_2 F_1 (-m, \b,\g,z) = \sum_{n=0}^m \frac{(\b)_n \Gamma(m+1) }{(\g)_n \G(m+1-n)} \frac{(-z)^n}{n!}
\ee
While taking derivaties with respect to $a$ and $c$ is straightforward,  the Fourier transform of the $\ln \bar z_{23}$ term involves differentiating with respect to $b$ and only then setting $b$ to be an integer. In the following, we explicitly evaluate the derivative of the hypergemetric with respect to a parameter, which only then is set to be an integer

%
%
\be
\left. \p_\a ({}_2 F_1 (\a, \b,\g,z))\right|_{a=-m} = \lim_{\e \r 0} \sum_{n=0}^\infty \frac{(\b)_n \Gamma(m+1) }{(\g)_n \G(m-\e+1-n)} \frac{(-z)^n}{n!} \left( \psi (n-m+\e)-\psi(-m+\e)\right)
\ee
where $\psi$ is the digamma function, which satisfies
\be
\psi (n-m+\e)-\psi(-m+\e) = \sum_{k=0}^{n-1} \frac{1}{k-m+\e}
\ee
The expansion on the right-hand side can thus be split into two sums, one for $n \leq m$, and one for $n>m$
\bea
RHS&=& \sum_{n=0}^m \frac{(\b)_n \Gamma(m+1) }{(\g)_n \G(m+1-n)} \frac{(-z)^n}{n!} \sum_{l=0}^{n-1} \frac{1}{l-m} +  \lim_{\e \r 0} \sum_{n=m+1}^\infty \frac{(\b)_n \Gamma(m+1) }{(\g)_n \G(m-\e+1-n)} \frac{(-z)^n}{n!} \sum_{l=0}^{n-1} \frac{1}{l-m+\e} \nonumber \\
&=&  \sum_{n=0}^m \frac{(\b)_n \Gamma(m+1) }{(\g)_n \G(m+1-n)} \frac{(-z)^n}{n!} \sum_{l=0}^{n-1} \frac{1}{l-m} + (-1)^m m!\, z^{m+1} \sum_{k=0}^\infty \frac{(\b)_{m+1+k} \, k!  }{(\g)_{m+1 + k}  } \frac{z^k}{(m+1+k)!} 
\eea
where only the terms with $l=m$ contribute in the latter sum. Letting $n=m+1 +k$ in the last term, the sum can be converted into a hypergeometric function, and equals

\be
z^{m+1} \frac{(-1)^m (\b)_{m+1}}{(m+1)\, (\g)_{m+1}} \, {}_3 F_2 (1,1,1+m+\b; m+2, 1+m +\g; z)
\ee
Notice that neither this, nor the finite sum in the first term,  exhibits a logarithmic behaviour.

\end{document}